\documentclass[final,1p,times]{elsarticle}
\usepackage{epsfig}
\usepackage{subcaption}
\journal{Physica A}
\usepackage{xcolor}

\DeclareMathAlphabet{\oldcal}{OMS}{cmsy}{m}{n}

\usepackage[english]{babel}
\usepackage[utf8x]{inputenc}
\usepackage[T1]{fontenc}
\usepackage{multirow}
\usepackage{tcolorbox}
\usepackage{longtable}
\usepackage{amsmath}
\usepackage{graphicx}
\usepackage[colorinlistoftodos]{todonotes}
\usepackage[colorlinks=true, allcolors=blue]{hyperref}
\usepackage{tikz-cd}

\usepackage{amsmath,amssymb,lmodern,bm,pxfonts}
\usepackage[mathscr]{euscript}
\let \eucal \mathscr
\usepackage{mathrsfs}
\DeclareMathAlphabet{\mathpzc}{OT1}{pzc}{m}{it}
\DeclareMathVersion{normal2}
\usepackage{float}
\def\hlinewd#1{%
\noalign{\ifnum0=`}\fi\hrule \@height #1 %
\futurelet\reserved@a\@xhline}
\usepackage{multicol}
\usepackage{array}
    \newcolumntype{P}[1]{>{\centering\arraybackslash}p{#1}}
    \newcolumntype{M}[1]{>{\centering\arraybackslash}m{#1}}

\begin{document}

\title{Model A of critical dynamics: 5-loop $\varepsilon$ expansion study}

\author[label1,label2]{L.\,Ts.\,Adzhemyan}
\author[label1]{D.A.\,Evdokimov}
\author[label2,label3,label4]{M.\,Hnati\v{c}}
\author[label5]{E.~V. Ivanova}
\author[label1,label2]{\corref{cor1}M.\,V.\,Kompaniets}
\ead{m.kompaniets@spbu.ru} 
\author[label6]{\corref{cor2}A.\,Kudlis}
\ead{andrew.kudlis@metalab.ifmo.ru} 
\cortext[cor1]{}
\author[label1]{D.V.\,Zakharov}

\address[label1]{Saint Petersburg State University, 7/9 Universitetskaya Embankment, St. Petersburg, 199034 Russia}

\address[label2]{Bogoliubov Laboratory of Theoretical Physics, Joint Institute for Nuclear Research, 6 Joliot-Curie, Dubna, Moscow region, 141980, Russian Federation}

\address[label3]{Department of Theoretical Physics, SAS, Institute of Experimental Physics, Watsonova 47, 040 01 Ko\v{s}ice, Slovak Republic}

\address[label4]{Pavol Jozef \v{S}af\'{a}rik University in Ko\v{s}ice (UPJ\v{S}), \v{S}rob\'{a}rova 2, 041 80 Ko\v{s}ice, Slovak Republic}

\address[label5]{New Jersey Institute of Technology, 323 Dr Martin Luther King Jr Blvd, Newark, NJ 07102, USA}

\address[label6]{Faculty of Physics, ITMO University, Kronverkskiy prospekt 49, St. Petersburg 197101, Russia}

\begin{keyword}
renormalization group, critical dynamics, multi-loop calculation, critical dynamic exponent z, $\varepsilon$ expansion
\end{keyword}

\begin{abstract}
We have calculated the five-loop RG expansions of the $n$-component A model of critical dynamics in dimensions $d=4-\varepsilon$ within the Minimal Subtraction scheme. This is made possible by using the advanced diagram reduction method and the Sector Decomposition technique adapted to the problems of critical dynamics. The $\varepsilon$ expansions for the critical dynamic exponent $z$ for an arbitrary value of the order parameter dimension $n$ are derived. Based on these series, the numerical estimates of $z$ for different universality classes are extracted and compared with the results obtained within different theoretical and experimental methods.
\end{abstract}

\maketitle
\section{Introduction}

The observable in the vicinity of critical point physics already for half a century continues to surprise with its peculiarities. The forgetfulness of the undergoing second-order phase transition physical systems regarding their microscopic features allows dividing such systems into so-called universality classes~\cite{green2017,nla.cat-vn4318275}. The systems united by one universality class should demonstrate the same critical behavior. Moreover, an elegant feature of the theory is the fact that each class of universality is determined only by the most fundamental characteristics: spatial dimensionality, the number of order parameter components, the range of interaction, as well as the symmetry of the system above the criticality~\cite{WILSON197475,RevModPhys.46.597,zjj1989,PELISSETTO2002549,V04}. However, universality is not a panacea, for example it does not apply to the value of critical temperature; it concerns only a limited set of physical observables, where critical exponents have been the most studied over the years. These, as a rule, non-integer numbers describe the power-law behavior of the thermodynamic characteristics near the continuous phase transition. Most of them are related to the consideration of a static problem. There is, however, an exponent that requires already to study the \textit{critical dynamics} of the system. It is the dynamic critical exponent $z$. The classification of various dynamic models was suggested in Ref.~\cite{HH77}; it is based on the conservation or not of both the energy of the system and the order parameter. The purely dissipative relaxational dynamics of the order parameter corresponds to the case of the so-called model A, where the critical exponent $z$ describes the critical slowing down connecting the correlation length $\xi$ and the typical time of fluctuations $\tau$ as:
\begin{equation}\label{eq:tau_sim_xi}
    \tau\propto\xi^z,
\end{equation} 
where $\xi$ and $\tau$ grow indefinitely when approaching the critical point. This behavior is caused by strong fluctuations that are present in the vicinity of the phase transition and due to its existence relation~\eqref{eq:tau_sim_xi} does not allow the system to achieve its equilibrium state. As for numerical calculations, the struggle is at the level of deviations of tenths and hundredths from the canonical value $z = 2$. 

In fact, the A model is the simplest dynamic generalization of the static $\phi^4$ field theory, that allows one to test various techniques being usual for static consideration of critical dynamics with the least possible difficulties. The treatment of this model can be realized by using different Monte Carlo (MC) methods, high-temperature (HT) expansion technique, as well as using various field theoretical (FT) approaches. Unfortunately, each method has its drawbacks. This can be either a finite lattice size or poor convergence of expansions. In this situation, in order to understand the true picture of the phenomenon, a comprehensive study of the problem using all available approaches is vitally important. Below, we focus mostly on the specific results that were obtained within the mentioned methods in the case of different values of the order parameter dimensionality $n$, as well as the spatial dimensions $d$ without detailed analysis of reasons of possible contradictions between them. In this work, we will associate each of the universality classes or A model for specific $d$ and $n$ with the symbol $\bm{\eucal{A}}^{d,n}$ while avoiding mentioning the Ising model or Heisenberg one in order to exclude possible misunderstanding regarding the relation with other dynamic models as the J or C models of critical dynamics. Note that most results are naturally devoted to the one-, two-, and three-component models A, which physically describe the critical slowing down in easy-axis, planar and isotropic ferromagnets, respectively. Among them, most attention of researchers was riveted to the $\bm{\eucal{A}}^{2,1}$ and $\bm{\eucal{A}}^{3,1}$ models as well as to the $\bm{\eucal{A}}^{3,3}$ one. 

Let us begin with MC analysis of lattice models, which can be realized by means of different options. The system can be prepared both at equilibrium or out of it. For example, one can suddenly change the temperature of the system to the critical one and allow it to relax towards equilibrium. In addition, one can slowly change the external field for a system located in the critical region. It should also be noted that in practice expression~\eqref{eq:tau_sim_xi} is replaced by the dependence on the lattice size $L$ instead of the infinite correlation length as: $\tau\propto L^z$. A comprehensive review of different nonequilibrium universality classes for lattice systems is given in Ref.~\cite{Ordor2004}. For almost forty years, a large number of works presented in which lattice calculations of the dynamic critical exponent $z$ of the $n$-component model A in two and three dimensions were performed~\cite{Wansleben1987,Wansleben1991,Mnkel1993,ito1993non3d,ito19932d,Matz1994,grassberger1995damage,Li1995,gropengiesser1995damage,Nightingale1996,stauffer1996flipping,Soares1997,Wang1997,Wang98,jaster1999short,godreche2000response,Ito2000,Nightingale2000,Lei2007,Murase2008,Collura_2010,hasenbusch2020,peczak1990monte,peczak1993,Fernandes_2006,Pospelov_2019,Ying2001XY,Astillero2019}. The corresponding numerical estimates of $z$ are collected in Table~\ref{tab:crit_exp_z} and in Table~\ref{tab:crit_exp_z_n} for the $n=1$ and $n>1$ cases, respectively. Each of the works deserves a separate consideration, but this is not the purpose of the present paper. Let us consider only a few of them trying to cover different methods, for the rest of them we give only numerical estimates of the exponent $z$.

As was said above, a significant number of works were dedicated to the $\bm{\eucal{A}}^{3,1}$ model. In Ref.~\cite{Wansleben1987}, the authors investigated the equilibrium dynamic critical behaviour of the one-component model on a simple cubic lattice, where by measuring the decay of the time-displaced correlation function of magnetization at the critical temperature, they estimated the critical exponent $z$ as $2.03(4)$. The authors emphasized that such an estimate agrees with the limitations imposed by the renormalization group approach: $z-2>0$. In Ref.~\cite{Mnkel1993}, the authors suddenly quenched the system from an ordered state at $T=0$ to a critical one. For three dimensions by means of both Glauber and Metropolis dynamics they found: $z=2.08(3)$. In Ref.~\cite{Matz1994}, the authors addressed to the so-called \textit{damage spreading} technique which consists in observing the dynamics and appropriate simultaneous changing of boundary spin layers of almost identical systems. Taking into account logarithmic corrections, the authors extracted the following estimates for $z$ in the $\bm{\eucal{A}}^{3,1}$ model case: $2.05(4)$. In Ref.~\cite{grassberger1995damage}, by means of the improved damage spreading method, the authors found the refined exponents values for 2d and 3d cases as $2.172(6)$ and $2.032(4)$, respectively. In Ref.~\cite{Nightingale1996}, in order to accurately determine the critical behaviour of the relaxation time, the authors introduced the \textit{variance-reducing} Monte Carlo algorithm. For the $\bm{\eucal{A}}^{2,1}$ they found $2.1665(12)$ for $z$ by means of the single-spin flip Markov dynamics accompanied by a probability determination according to the heat-bath method. In other works, the authors analyzed \textit{short time dynamics} of the system~\cite{Soares1997}. It turned out that short time behaviour does not depend on the initial condition. The authors managed to extract the value of the critical dynamic exponent $z$ from observing relaxation to equilibrium of well-chosen variables. The exponents were estimated as $2.16(3)$ and $2.075(35)$ for 2d and 3d cases, respectively. In Ref.~\cite{Wang98}, the nonequilibrium relaxation was studied, whereas the initial state is chosen as a completely ordered  state at the critical temperature. By computing the magnetization, energy per bond, and three-spin correlation, the authors found the following estimate for the $\bm{\eucal{A}}^{2,1}$ model: $z=2.169(3)$. Finally, let us pay special attention to the recent calculations~\cite{hasenbusch2020} that partly prompted us to implement the present work. There, the critical exponent $z$ for the $\bm{\eucal{A}}^{3,1}$ model was estimated as: $2.0245(15)$. This result was obtained by considering the so-called improved Blume-Capel model on a simple cubic lattice that allows authors to eliminate the leading corrections to scaling. The work also noted that the difference in numerical estimates obtained within previous MC simulations with the results given by, for example, FT methods, which is not covered by the error bars, is due to incorrect data processing, i.e., neglect of such scaling corrections.

The MC simulations for the $\bm{\eucal{A}}^{3,3}$ model are usually performed on a simple cubic lattice. In Refs.~\cite{peczak1990monte,peczak1993}, the authors addressed to the equilibrium dynamic critical behaviour of the autocorrelation function in order to analyze the dependence of the relaxation time on lattice size and find the corresponding numerical estimate for the exponent $z$ as $1.96(6)$. The number was found in good agreement with the result obtained by means of the above mentioned short time MC simulations $1.976(9)$ in Ref.~\cite{Fernandes_2006}. These estimates, however, strongly contradict the renormalization group basis. This statement was recently supported in~\cite{Pospelov_2019}, where the authors managed to obtain the following value $2.035(4)$ by analyzing the evolution from  various  initial  states. In Ref.~\cite{Astillero2019}, the authors attacked the problem both in and out of equilibrium states. There, by computing the autocorrelation time at equilibrium, they extracted $z=2.033(5)$, while in the out-of-equilibrium regime, they found the following estimate: $z=2.04(2)$. 

As already mentioned, less attention was paid to the rest of the universality classes. In particular, in Ref.~\cite{Ying2001XY}, for the $\bm{\eucal{A}}^{2,2}$ model the authors used MC simulations to investigate the short time behavior of the system dynamics starting from both ordered and disordered states. They estimated the exponent $z$ as $2.04(1)$. In Ref.~\cite{Zheng2003XY}, the authors obtained the whole set of estimates. Due to the relatively low accuracy, unfortunately, they all were roughly estimated as $2$.

As for the FT calculations, there is also a vast variety of  results~\cite{Folk_2006}. However, from the very beginning, it should be noted that in contrast to static problems, the degree of elaboration of their dynamic counterparts, due to the advancement of perturbation theory (PT) to higher orders, leaves much to be desired. Moreover, as is known, there are a number of different renormalization group (RG) approaches. Let us review them consequentially. 

First, the RG analysis can be performed in fixed spatial dimensionality. The record-high result here is the four-loop analysis performed in Ref.~\cite{Prudnikov1997}. For $\bm{\eucal{A}}^{2,1}$ and $\bm{\eucal{A}}^{3,1}$ models for $z$ the authors obtained $2.093$ and $2.017$, respectively. However, these numbers were extracted using the  most straightforward resummation technique -- the method of Pad\'e approximants. Later, however, these estimates were shifted, which was made possible by using the advanced resummation strategies~\cite{krinitsyn2006calculations}. The improved scheme allowed the authors to extract the following numbers: $2.0842(39)$ and $2.0237(55)$ for 2d and 3d cases, respectively.  
\begin{table}[t!]
 \centering
    \caption{Numerical estimates of the dynamic critical exponent $z$ for the $\bm{\eucal{A}}^{2,1}$ and $\bm{\eucal{A}}^{3,1}$ models obtained by means of different theoretical and experimental approaches. Abbreviations: ED --  equilibrium dynamic critical behaviour; SQtCT -- sudden quench to critical temperature; DS -- damage spreading; EVm -- eigenvalue method; STD -- short time dynamics; NED -- nonequilibrium dynamics; IBC -- improved Blume-Capel model; PR -- Pad\'e-resummed expansions; FSD -- fixed spatial dimensionality; NP -- nonperturbative; wor -- without regulators; r$_i$ -- different regulators. Subscript $f$ denotes "face centred cubic" lattice, while $b$ -- body centred cubic one.} 
    \label{tab:crit_exp_z}
     \setlength{\tabcolsep}{20.07pt}
    \begin{tabular}{llllll}
      \hline
      \hline
      Method & Ref. & Year &2d & 3d  \\
      \hline
        MC: ED & \cite{Wansleben1987}  & 1987 & - & $2.03(4)$  \\
        MC: ED & \cite{Wansleben1991}  & 1991 & - & $2.04(3)$  \\
        MC: SQtCT & \cite{Mnkel1993}  & 1993 & $2.21(3)$ & $2.08(3)$  \\
        MC: SQtCT & \cite{ito1993non3d}  & 1993 & - & $2.073(16)$  \\
        MC: SQtCT & \cite{ito19932d}  & 1993 & 2.165(10) & $2.073(16)$  \\
        MC: DS & \cite{Matz1994}  & 1994 & - & $2.05(4)$  \\
        MC: DS & \cite{grassberger1995damage}  & 1995 & $2.172(6)$ & $2.032(4)$  \\
        MC: DS & \cite{Li1995}  & 1995 & $2.1337(41)$ & -  \\
        MC: DS & \cite{gropengiesser1995damage}  & 1995 & $2.18(2)$ & $2.04(1)$  \\
        MC: EVm & \cite{Nightingale1996}  & 1996 & $2.1665(12)$ & -  \\
        MC: STD & \cite{Soares1997}  & 1997 & $2.16(3)$ & $2.075(35)$  \\
        MC: DS & \cite{Wang1997}  & 1997 & $2.166(7)$ & -  \\
        MC: NED & \cite{Wang98}  & 1997 & $2.169(3)$ & -  \\
        MC: STD & \cite{jaster1999short}  & 1999 & - & $2.042(6)$  \\
        MC: SQtCT & \cite{godreche2000response}  & 2000 & $2.17$ & -  \\
        MC: SQtCT & \cite{Ito2000}  & 2000 & - & $2.055(10)$  \\
        MC: EVm & \cite{Nightingale2000}  & 2000 & $2.1667(5)$ & -  \\
        MC: NED & \cite{Lei2007}  & 2007 & $2.16$ & -  \\
        MC: SQtCT & \cite{Murase2008}  & 2007 & $2.165(15)$ & $2.065(25)_{b}$ \\
        MC: SQtCT & \cite{Murase2008}  & 2007 & - & $2.057(25)_f$ \\ 
        MC: IBC & \cite{Collura_2010}  & 2010 & - & $2.020(8)$  \\
        MC: IBC  & \cite{hasenbusch2020}  & 2019 & - & $2.0245(15)$  \\
        HT: PR & \cite{Racz1976HT}  & 1976 & $2.125(1)$ & -  \\
        HT: PR & \cite{dammann1993HT}  & 1993 & $2.183(5)$ & -  \\
        RG: FSD & \cite{Oerding_1995} 
        & 1995 & $2.124$ & $2.022$\\
        RG: FSD & \cite{Prudnikov1997}  & 1997 & $2.093$ & $2.017$  \\
        RG: FSD & \cite{krinitsyn2006calculations}  & 2006 & $2.0842(39)$ & $2.0237(55)$  \\
        RG: NP & \cite{Canet_2007}  & 2007 &  $2.16(1)$ & $2.09(4)$  \\
        RG: NP & \cite{PhysRevD.92.076001}  & 2015 & - & $2.025$  \\
        RG: NP(wor) & \cite{DuclutDelamotte2017} & 2017 & $2.28$ & $2.032$  \\
        RG: NP(r$_1$) & \cite{DuclutDelamotte2017}  & 2017 & $2.16$ & $2.024$  \\
        RG: NP(r$_2$) & \cite{DuclutDelamotte2017}  & 2017 & $2.15$ & $2.024$  \\
        RG: NP(r$_3$) & \cite{DuclutDelamotte2017}  & 2017 & $2.14$ & $2.023$  \\
        Exp:  & \cite{NG15}&2015& -& $2.06$\\
    \hline
    \hline
    \end{tabular}
\end{table}
\begin{table}[t!]
 \centering
    \caption{Numerical estimates of the dynamic critical exponent $z$ for the $\bm{\eucal{A}}^{2,n}$ and $\bm{\eucal{A}}^{3,n}$ models obtained by means of different theoretical approaches. Most cases relate to the planar ($n=2$) and Heisenberg ($n=3$) models. Abbreviations: INS -- inelastic neutron-scattering.}
    \label{tab:crit_exp_z_n}
     \setlength{\tabcolsep}{12.8pt}
    \begin{tabular}{lllllll}
      \hline
      \hline 
      Method & Ref. & Year & Num. of. comp.&2d & 3d  \\
      \hline
        MC: STD  & \cite{Ying2001XY} & 2001 & $n=2$ & $2.04(1)$ & \\
        RG: NP(wor)  & \cite{DuclutDelamotte2017}  & 2019 & $n=2$ & - & $2.029$  \\
        RG: NP(r$_1$)  & \cite{DuclutDelamotte2017}  & 2019 & $n=2$ & - & $2.024$  \\
        RG: NP(r$_2$)  & \cite{DuclutDelamotte2017}  & 2019 & $n=2$ & - & $2.023$  \\
        MC: ED  & \cite{peczak1990monte}  & 1990 & $n=3$ & - & $1.96(7)$  \\
        MC: ED  & \cite{peczak1993}  & 1993 & $n=3$ & - & $1.96(6)$  \\ 
        MC: STD  & \cite{Fernandes_2006}  & 2006 & $n=3$ & - & $1.976(9)$  \\                
        MC: NED & \cite{Pospelov_2019}  & 2019 & $n=3$ & - & $2.035(4)$  \\                
        MC: ED  & \cite{Astillero2019}  & 2019 & $n=3$ & - & $2.033(5)$  \\
        MC: NED  & \cite{Astillero2019}  & 2019 & $n=3$ & - & $2.04(2)$  \\
        RG: NP(wor)  & \cite{DuclutDelamotte2017}  & 2019 & $n=3$ & - & $2.025$  \\
        RG: NP(r1)  & \cite{DuclutDelamotte2017}  & 2019 & $n=3$ & - & $2.021$  \\
        RG: NP(r2)  & \cite{DuclutDelamotte2017}  & 2019 & $n=3$ & - & $2.021$  \\
        Exp: ESR & \cite{Dunlap1980} & 1980 & $n=3$ & - &  $2.04(7)$ \\
        Exp: INS & \cite{Bohn1984}   & 1984 & $n=3$ & - &  $2.09(6)$\\  
    \hline
    \hline
    \end{tabular}
\end{table}

An alternative approach also applied to the problem is the so-called nonperturbative or functional renormalization group (NPRG)~ \cite{Canet_2007,PhysRevD.92.076001,DuclutDelamotte2017}. In Ref.~\cite{Canet_2007}, the authors obtained the following estimates for the two $2.16(1)$ and three $2.09(4)$ dimensions. A noticeably different number from the latter for $\bm{\eucal{A}}^{3,1}$ model was found in Ref.~\cite{PhysRevD.92.076001}: $2.025$. The authors of Ref.~\cite{DuclutDelamotte2017} suggested the whole set of numerical estimates which correspond to the application of different NPRG regulators and without them. All numbers have been added to Table~\ref{tab:crit_exp_z}. 

Consider now the field theory approach that revolutionized the critical behavior theory by its appearance. The Wilson idea to take as a formal small parameter the difference between the upper critical dimension, above which the Landau theory is workable, and the physical value of the system spatial dimension allowed theoreticians to find anomalous deviations of different physical observables from their canonical values at criticality. Despite the asymptotic nature of the $\varepsilon$ expansions, even taking into account the first two or three terms is sufficient to give adequate numerical estimates for the main mass of physically interesting quantities. In the present work, the dynamic critical exponent $z$ is obtained within this formalism. The two-loop result was found almost fifty years ago in Ref.~\cite{Halperin_1972}. Ten years later, three-loop results were found, which were record-high for a long time~\cite{Antonov1984}. The first attempt to calculate four-loop expansions was made in Ref.~\cite{ANS08} with subsequent resummation in Ref.~\cite{NSS9}. Only ten years later, however, the accuracy of four-loop calculations was notably improved~\cite{AIKV18} using the new \textit{diagram reduction technique}. This method can significantly decrease the number of diagrams and make it possible to carry out calculations in real time. Thanks to this technique, we are able to calculate the five-loop contribution to the dynamic critical exponent $z$ within $d=4-\varepsilon$ dimension for an arbitrary number of order parameter components. In such a high order of PT, the reduction in the number of diagrams turned out to be crucial: from $1025$ to $201$. The expansion of this length in combination with the novel resummation approach (we call it the \textit{free boundary condition method}, which was developed based on the \textit{boundary condition method} suggested by R. Guida and J. Zinn-Justin), makes it possible to extract highly accurate estimates for $z$ for different values of spatial and spin dimensions. This paper is an extended version of the recently published letter (ref. on our future paper), where details of numerical calculations were omitted, and numerical estimates were given for the most significant case -- $\bm{\eucal{A}}^{3,1}$ model. 

Before we start with specific calculations, let us review some experimental results relevant to the problem. There is an extensive list of physical systems that are described by the model A. For example, the recent results obtained in Refs.~\cite{HONKONEN2019105,Zhavoronkov2019} have demonstrated that the phase transition into a superfluid state is also described by the model A. In order to extract critical behavior characteristics, one can address the measurement of thermal transport properties. In particular, special attention of experimentalists is attracted by thermal diffusivity. It is worth noting that this type of measurement is challenging because, on the one hand, one needs to create temperature gradients; on the other, it must be extremely small in order to preserve the system at criticality. For instance, in Ref.~\cite{MMB95}, the authors analyzed the behaviour of strongly anisotropic antiferromagnet FeF$_2$ at antiferromagnetic paramagnetic transition. The work contains debates on the applicability of different dynamic models~(A or C) to the system. The choice in favor of the model C is only  qualitative, justified only by weak damping of energy, only with noticeable damping of the order parameter. In Ref.~\cite{NG15}, the magnetoelectric dynamics in the multiferroic chiral antiferromagnet MnWO$_4$ was analyzed. The observation of critical slowing down of magnetoelectric fluctuations allowed the authors to conclude about the validity of the theoretical description of this physical system within the model with overdamped magnetic 3d-Ising order parameter, i.e., the $\bm{\eucal{A}}^{3,1}$ model. The authors of other works~\cite{LF15,AuAgZn22018} studied the continuous phase transition of the AuAgZn$_2$ alloy by coherent x-ray scattering. Having quenched the alloy samples, they observed the motion of interfaces between ordered domains. Using the critical system behaviour, they also came to the conclusion that the critical dynamics of such a system corresponds to the A model, although the number for the dynamic critical exponent is $1.96(11)$. This number indicates that the agreement between theory and experiments is only qualitative due to the error bar of experimental numbers. Similar results were obtained in Ref.~\cite{LB2002}. In Ref.~\cite{Hohenemser1982}, different ferromagnets were investigated by means of electron-spin-resonance and hyperfine interaction. However, the main purpose of the work was to connect the behavior of the materials with one or other dynamic models. In particular, the authors aimed to distinguish the physical systems described by the J model, where the approximate value of $z$ is $2.5$, from systems where the order parameter is not conserved, i.e., the A model of critical dynamics, for which the exponent $z$ can be roughly estimated as $2$. There was no talk of extracting some corrections as tenths or even hundredths to $z=2$. In Ref.\cite{Dunlap1980}, the authors measured the spin-relaxation time for the isotropic ferromagnet EuO above the critical temperature by means of the zero-field electron-spin resonance. They estimated the dynamic exponent $z$ as $2.04(7)$. The authors of different works aimed to analyze spin dynamics of EuS using inelastic neutron-scattering technique~\cite{Bohn1984}. For the critical exponent $z$ they obtained $2.09(6)$. In addition, some experiments by means of M\"{o}ssbauer spectroscopy were carried out on impurity systems, where the primary purpose was to determine whether the addition of impurities can change the class of universality or not~\cite{Rosov1992}.

The paper is organized as follows. In Sec.~\ref{Sec_description} the field-theoretical model and its renormalization are described. Then, in Sec.~\ref{Sec_reduction} the features of the diagram reduction technique are explained. Next, in Sec.~\ref{Sec_expansion}, the RG expansions for anomalous dimensions and critical exponents $z$ are found up to $O(\varepsilon^5)$ terms. In Sec.~\ref{Sec_resummation}, the five-loop numerical estimates for $z$ in case of different $\bm{\eucal{A}}^{d,n}$ models are presented. At the end, some conclusion will be drawn.

\section{Description of the model} \label{Sec_description}
The nonrenormalized action of the model A of critical dynamics is defined by a set of two non-renormalized $n$-component fields $\phi_0\equiv \{\psi_0,\,\psi'_0\}$ and can be expressed in the form \cite{V04}:
\begin{equation}
S_0(\phi_0)
=\lambda_0 \psi_0' \psi_0' + \psi_0'[ -\partial_t \psi_0 + \lambda_0 (\partial^2 \psi_0 - m_0 \psi_0 - \frac{1}{3!}g_0\psi_0^3)]\, ,
\end{equation}
where $\lambda_0$ is the Onsager coefficient and $g_0$ is the coupling constant.
The model A is multiplicatively renormalizable. The renormalized action in the spatial dimensionality $ d = 4- \varepsilon $ reads as follows:
\begin{equation}\label{SR}
S_{\text{R}}=Z_1 \lambda \psi' \psi' + \psi'[-Z_2 \partial_t \psi + \lambda (Z_3 \partial^2 \psi - Z_4 m \psi - \frac{1}{3!}Z_5 \mu^{\varepsilon}g \psi^3)]\,,
\end{equation}
where the renormalized fields and parameters are expressed in terms of bare ones as:
\begin{equation}\label{ren}
\lambda_0=\lambda Z_{\lambda},\quad m_0=m Z_{m},\quad g_0=g \mu^{\varepsilon} Z_g,\quad \psi_0=\psi Z_{\psi}, \quad \psi_0'=\psi'Z_{\psi'}\,.
\end{equation}
The renormalization constants $Z_i$ are associated with the renormalization constants from \eqref{SR} by the following relations:
\begin{eqnarray}\label{ren1}
& Z_1=Z_{\lambda}Z_{\psi'}^2\,, \quad Z_2=Z_{\psi'}Z_{\psi}\,, \quad Z_3=Z_{\psi'}Z_{\lambda}Z_{\psi}\,,& \\ \nonumber
&Z_4=Z_{\psi'}Z_{\lambda}Z_{m}Z_{\psi}\,, \quad Z_5=Z_{\psi'}Z_{\lambda}Z_g Z_{\psi}^3\,.&
\end{eqnarray}
Due to the fact that the renormalization constants $Z_\psi$, $Z_m$, and $Z_g$ coincide with their static counterparts, the renormalization constants of the $\phi^4$ model~\cite{V04}
\begin{equation}
Z_{\psi}=(Z_{\psi})_{st}\,,\quad Z_m=(Z_m)_{st} \,,\quad Z_g=(Z_g)_{st}\,,
\end{equation}
and due to the existence of the relation $Z_{\psi'}Z_\lambda=Z_\psi$~\cite{V04}, we can conclude that, in fact, we are only interested in $Z_{\lambda}$, which is characteristic of this dynamic model:
\begin{equation}\label{Zlambda}
Z_\lambda=Z_1 Z_{\psi'}^{-2}=Z_1^{-1}Z_\psi^2=Z_2^{-1}Z_\psi^2\,.
\end{equation}
For our purposes, it is convenient to calculate it using the renormalization constant $ Z_1 $, which is determined from the diagrams of the one-irreducible function $\Gamma_{\psi'\psi'}=\langle \psi'\psi'\rangle_{\text{1-irr}}/(2\lambda)$ at zero external frequency $ \omega $ with $ m = 0 $. For the PT expansion of this function, the charge $u=g  S_d/(2\pi)^d $ is used, where $S_d=2\pi^{d/2}/\Gamma(d/2)$ is the area of the $ d $-dimensional sphere of the unit radius. This expansion has the form:
\begin{equation}\label{Gamma}
\Gamma_{\psi'\psi'}|_{\omega=0, m=0}=Z_1(1+u^2 Z_g^2(\mu/p)^{2\varepsilon}A^{(2)}+u^3 Z_g^3(\mu/p)^{3\varepsilon}A^{(3)}+...)=Z_1(1+\sum_{i = 2}^{\infty} u^i Z_g^i(\mu/p)^{i\varepsilon}A^{(i)})\, .
\end{equation}
As the renormalization scheme we use the minimal subtraction one, in which the counterterms subtract only the polar contributions of $\varepsilon$ from the diagrams:
\begin{equation}\label{Z1}
  Z_1=1+\sum_{n \geq 1}Z_1^{(i)}(u)\,\varepsilon^{-i}\, .
\end{equation}

The coefficients $Z_1^{(i)}(u)$ are calculated as series in $u$ based on the requirements of absence of poles in $\varepsilon$ for $\Gamma_{\psi'\psi'}$ from~\eqref{Gamma}. The general theory states that the coefficients at the highest poles in~\eqref{Gamma} are expressed in terms of the coefficients at the first pole in such a way that the possible pole contributions at logarithms $(\log(\mu/p))^i$ should be reduced. This is a check for calculations of the coefficients $ A^{(i)}$ from Eq.~\eqref{Gamma}.

\section{Diagrammatic technique: reduction scheme} \label{Sec_reduction}
The coefficients $A^{(i)}$ were calculated using the diagram reduction procedure proposed in \cite{AIKV18}. Let us consider it using the five-loop diagrams with topology ''$e112|33|e44|44||$'' as an example (according to the Nickel notation). There are two different graphs for model A:
\begin{eqnarray}\label{primer}
\frac{1}{2}
\begin{matrix}
\includegraphics[angle=0, width=0.2\textwidth]{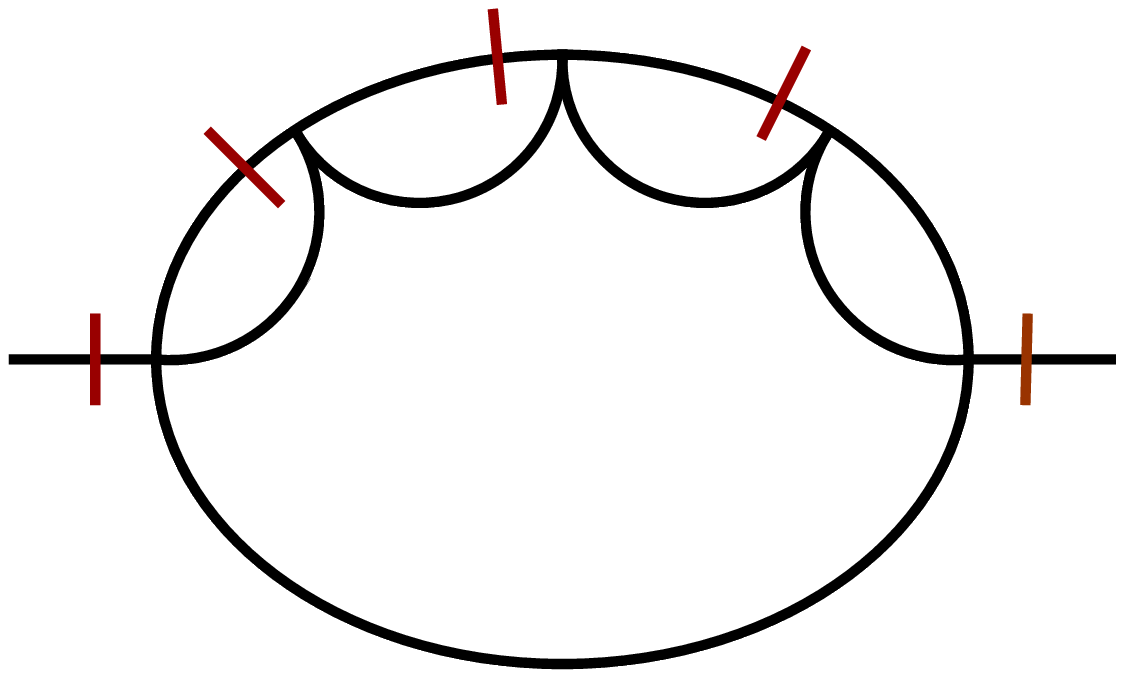}
\end{matrix}
+
\frac{1}{2} 
\begin{matrix}\label{5loop}
\includegraphics[angle=0, width=0.2\textwidth]{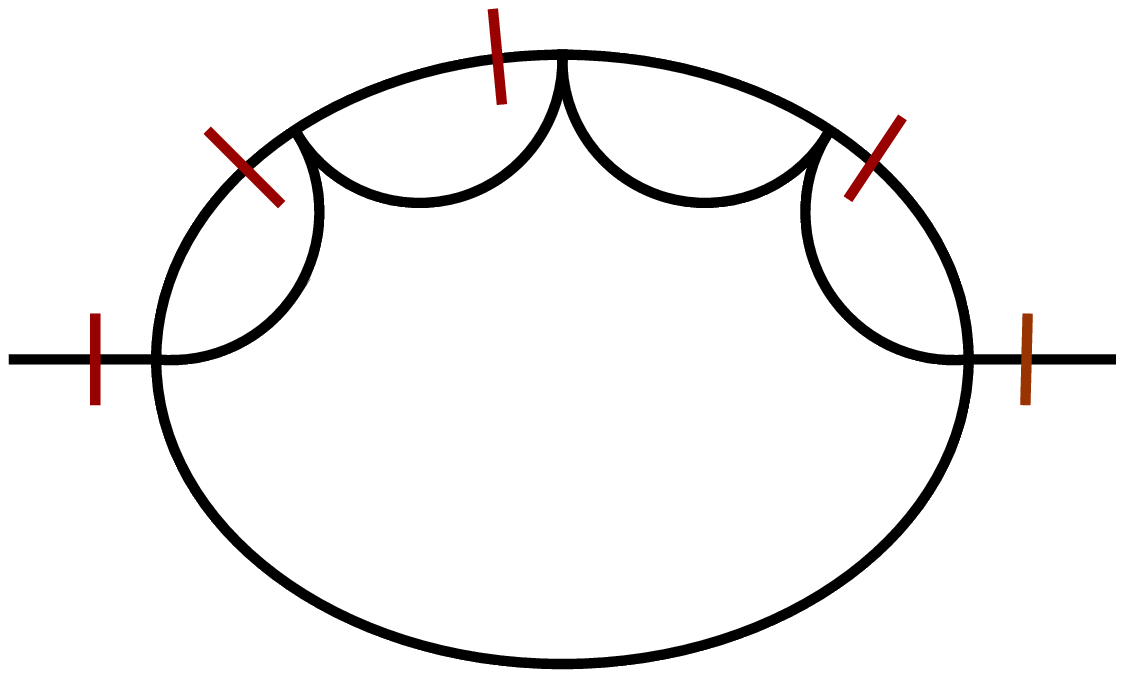}
\end{matrix}
\end{eqnarray}
Here the lines in the momentum-time ($k,\,t$) representation are matched with the following expressions:
\begin{eqnarray}
&&\begin{matrix}
\includegraphics[angle=0, width=0.1\textwidth]{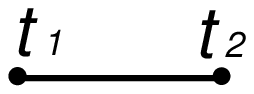}
\end{matrix}
 = \langle\psi(t_1)\psi (t_2) \rangle = \frac{1}{k^2} \exp^{-\lambda k^2|t_1-t_2|} \,, \\ 
&&\begin{matrix}
\includegraphics[angle=0, width=0.1\textwidth]{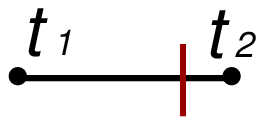}
\end{matrix}
 = \langle\psi(t_1)\psi' (t_2) \rangle = \theta(t_1-t_2)\exp^{-\lambda k^2(t_1-t_2)}\,.
\end{eqnarray}
The multiplier in front of the diagram is determined by its symmetry and the additional factor $1/2$ in the definition of the one-irreducible Green function $\Gamma_{\psi'\psi'}=\langle\psi'\psi'\rangle_{\text{1-irr}}/2\lambda$. The diagrams from Eq.~\eqref{primer} are matched with a set of ``time versions'' which correspond to all possible time ordering of vertices, taking into account $\theta$-functions within the lines $\langle\psi\psi'\rangle$. 
In each variant, the integration over time is easily performed. It turns out that the result of integration over time is the sum of contributions that already contain only integration over momentum space. These contributions can also be represented in the form of diagrams, in the case of diagrams from Eq.~\eqref{primer} the number of these diagrams is equal to 15.
In Ref.~\cite{AIKV18}, a method was proposed for grouping different time options of diagrams, significantly reducing their total number and simplifying the integrand expressions. Thus, the sum of the time versions of the diagrams from Eq.~\eqref{5loop} can be reduced to the sum of the following four unique diagrams:
\begin{eqnarray}\label{primer1}
\frac{1}{16}
\begin{matrix}
 \includegraphics[angle=0, width=0.15\textwidth]{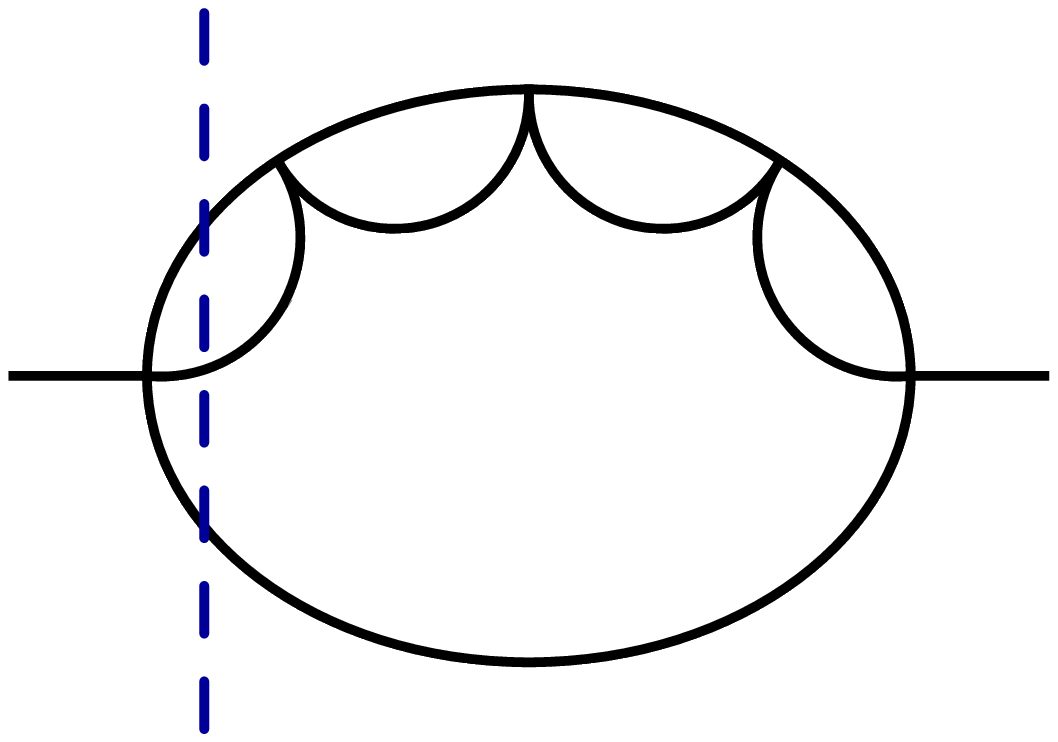}
\end{matrix}
+
\frac{1}{8}
\begin{matrix}
 \includegraphics[angle=0, width=0.15\textwidth]{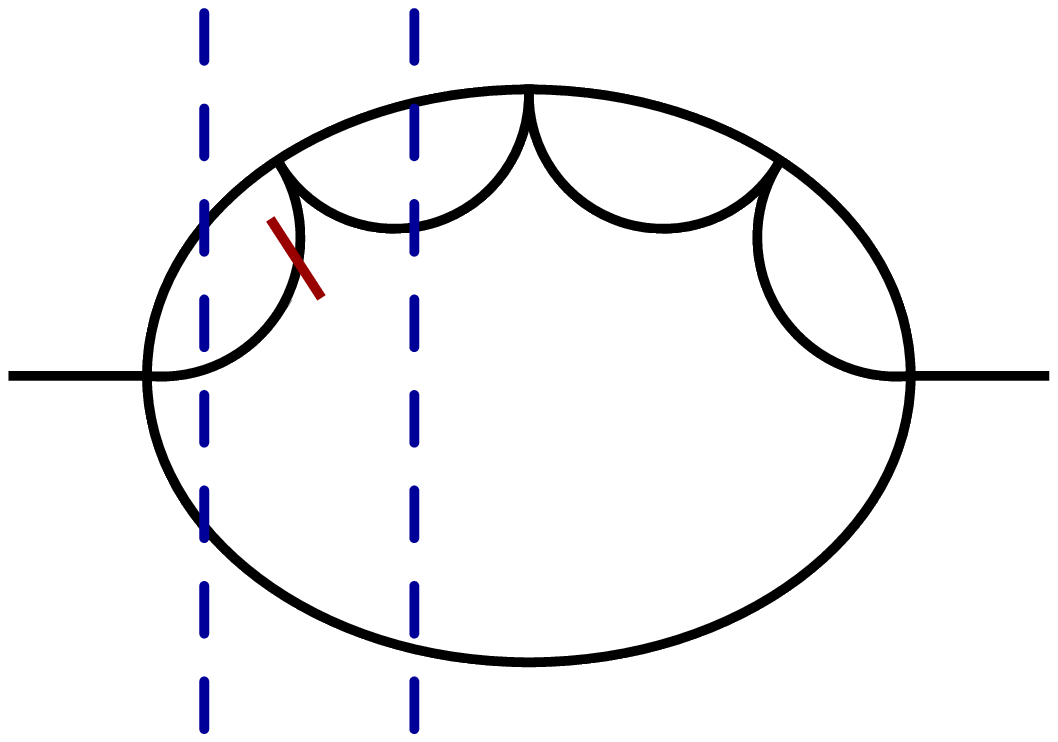}
\end{matrix}
+
\frac{1}{4}
\begin{matrix}
 \includegraphics[angle=0, width=0.15\textwidth]{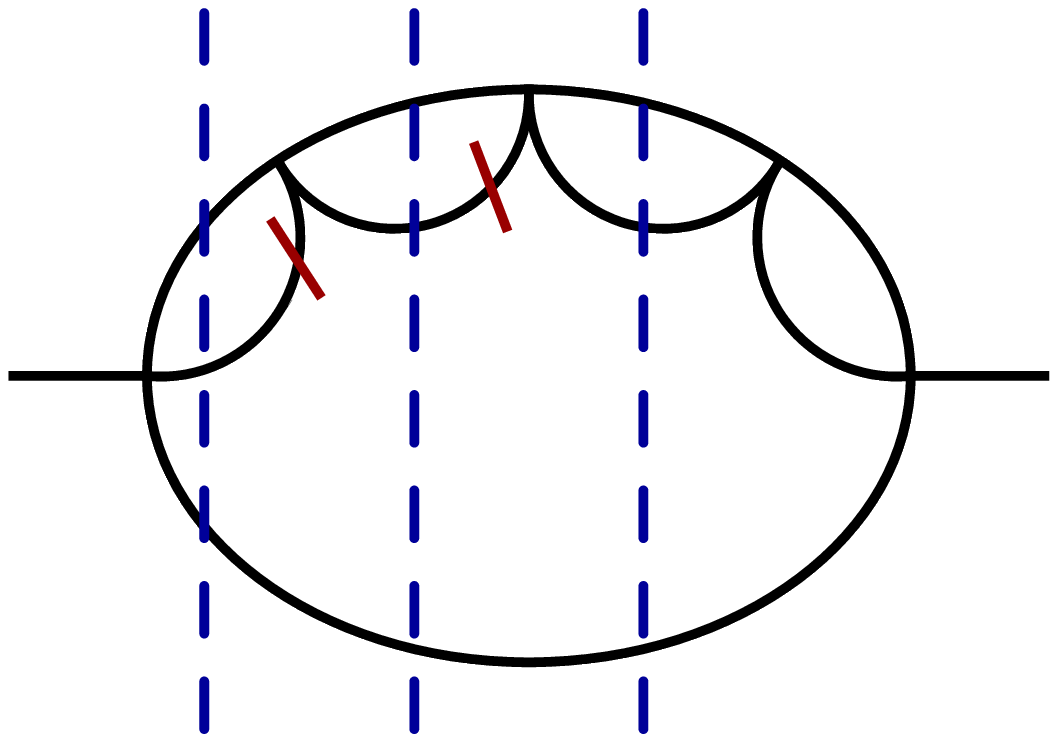}
\end{matrix}
+
\frac{1}{2}
\begin{matrix}
 \includegraphics[angle=0, width=0.15\textwidth]{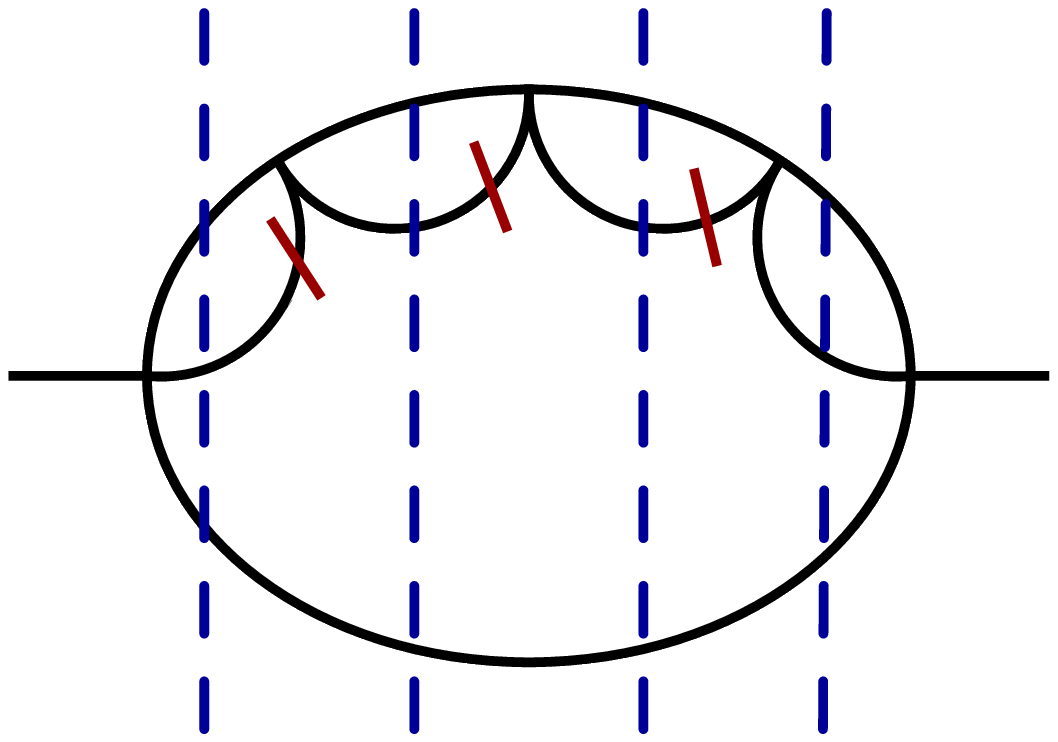}
\end{matrix}
\end{eqnarray}
In the reduced diagrams from Eq.~\eqref{primer1}, the lines with momentum $k$ without strikethrough are associated with the factor $1/k^2$, lines with strikethrough -- $1$, and the dashed lines -- fractions $1/S$, where $S$ is the sum of squares of the line momentum that the dashed line crosses. Thus, in the considered example, the number of integrals necessary for the calculation was reduced from 15 (the number of time versions) to 4. The factors near the reduced diagrams from Eq.~\eqref{primer1} correspond to the standard symmetry factors taking into account the additional factor $1/2$ in the definition of the one-irreducible Green function $\Gamma_{\psi'\psi'}=\langle \psi'\psi'\rangle_{\text{1-irr}}/(2\lambda)$. In Ref.~\cite{AIKV18}, the authors describe a general way of obtaining the reduced structures directly from the corresponding static ones.  
\noindent In Table~\ref{tab:topology}, we present all the necessary static diagrams up to the five-loop approximation, show the Nickel index for each of the static diagrams, the factors $a(n)$ which allow one to move from the one-component theory to an arbitrary number of order components, the number of the corresponding dynamic diagrams, the number of their corresponding time versions, and the number of diagrams after reduction. 
\renewcommand*{\arraystretch}{0.9}
\renewcommand*{\tabcolsep}{1.55pt}
\begin{longtable}{>{\centering\arraybackslash} m{1cm} 
>{\centering\arraybackslash} m{2cm}
ccccc}
\caption[Topology of all necessary static diagrams up to 5-loop contributions. Abbreviations: NDD -- number of dynamic diagrams; NTV -- number of the corresponding time versions; NRD -- number of reduced diagrams. Notation for topology of diagrams are chosen in accordance with Nickel notation.]{Topology of all necessary static diagrams up to 5-loop contributions. Abbreviations: NDD -- number of dynamic diagrams; NTV -- number of the corresponding time versions; NRD -- number of reduced diagrams. Notation for topology of diagrams is chosen in accordance with the Nickel notation.} 
\label{tab:topology} \\
\hline
\hline
\multicolumn{1}{c}{Num.} &
\multicolumn{1}{c}{Graph} &
\multicolumn{1}{c}{Topology} &
\multicolumn{1}{c}{$a(n)$} &
\multicolumn{1}{c}{NDD} &
\multicolumn{1}{c}{NTV} &
\multicolumn{1}{c}{NRD} \\
\hline
\endfirsthead
\multicolumn{7}{c}%
{\tablename\ \thetable{} Topology of all necessary static diagrams up to 5-loop contributions.} \\
\hline
\multicolumn{1}{|c}{Num.} &
\multicolumn{1}{|c}{Graph} &
\multicolumn{1}{|c}{Topology} &
\multicolumn{1}{|c}{$a(n)$} &
\multicolumn{1}{|c}{NDD} &
\multicolumn{1}{|c|}{NTV} &
\multicolumn{1}{|c}{NRD} \\
\hline
\endhead
\hline \multicolumn{7}{|r|}{{Continuation on the next page}} \\ \hline
\endfoot
\hline \hline
\endlastfoot
\multicolumn{7}{c}{2 loop}   \\ \hline
1  & \includegraphics[angle=0, width=0.1\textwidth]{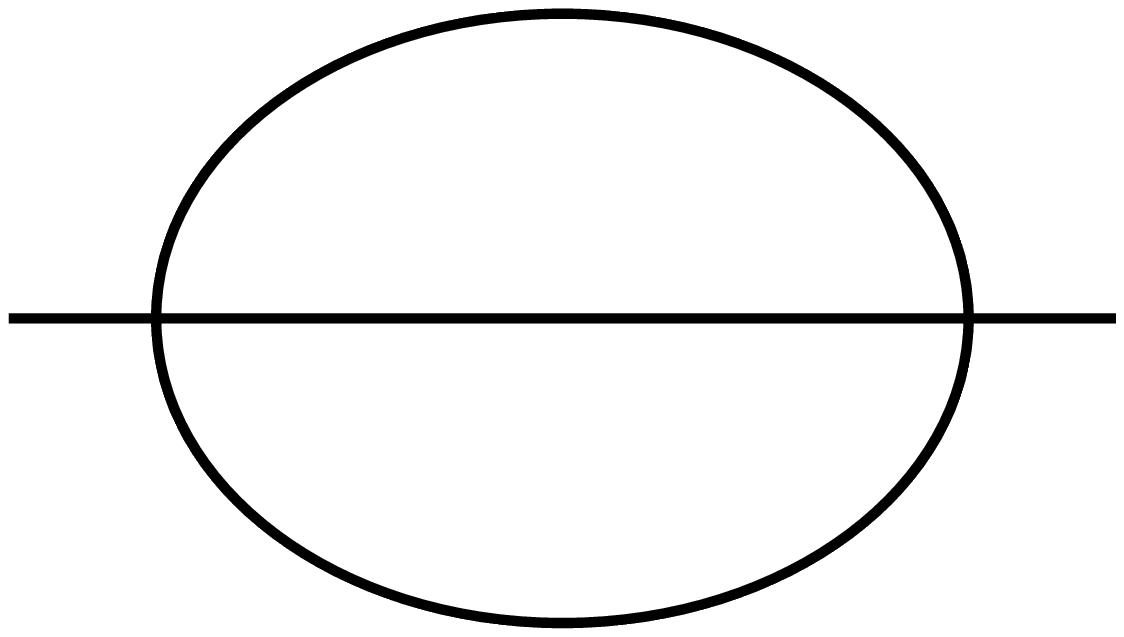} 
    & e111|e| & $(2+n)/3$& 1 & 2  & 1  \\ 
\multicolumn{4}{r}{Total:}&1 &  2  & 1 \\
\hline 
\multicolumn{7}{c}{3 loop}   \\ \hline 
1 & $ \includegraphics[angle=0, width=0.1\textwidth]{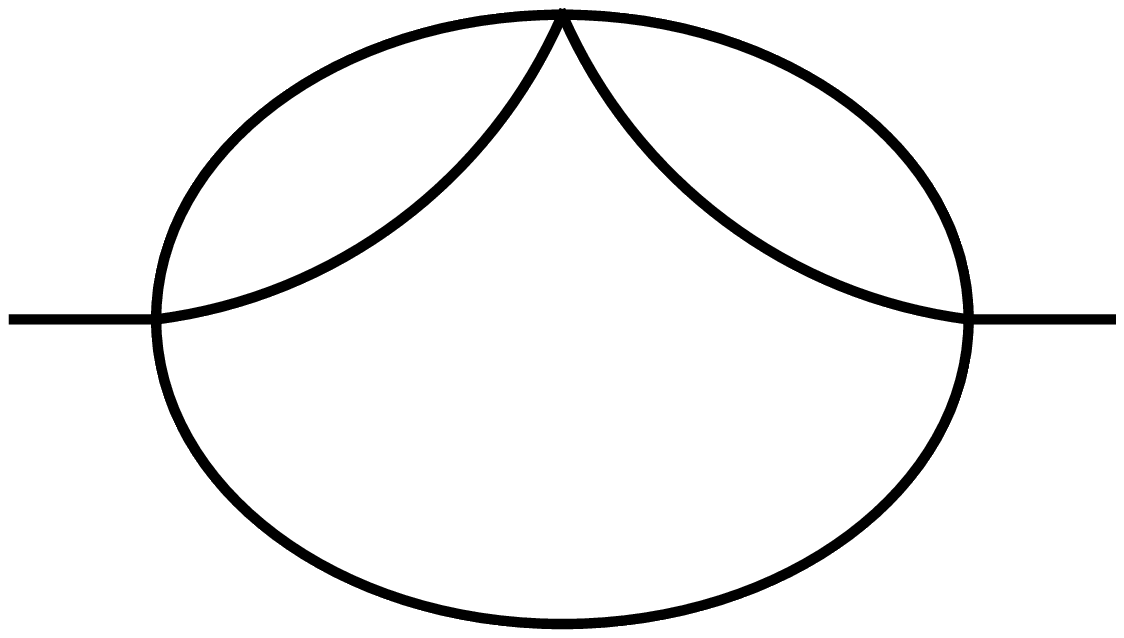}$ 
    & e112|22|e|&$(8+n)(2+n)/27$ & 1 & 2  & 2  \\
\multicolumn{4}{r}{Total:} & 1 & 2  & 2 \\
\hline
\multicolumn{7}{c}{4 loop}   \\ \hline 
1 & $ \includegraphics[angle=0, width=0.1\textwidth]{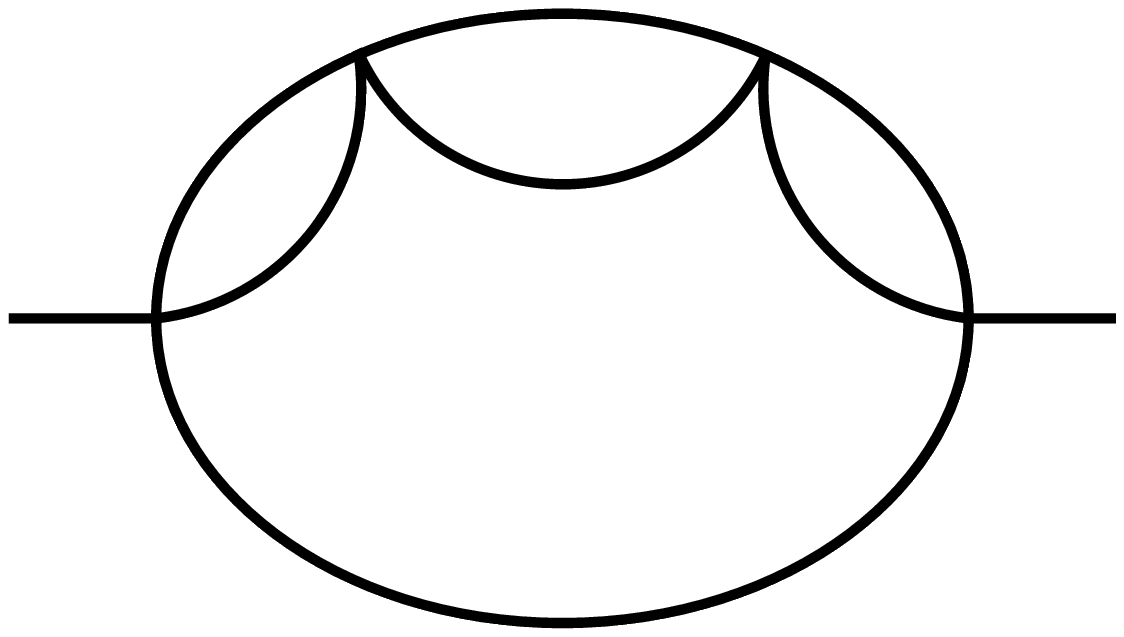}$ 
    & e112|33|e33||&$(n^2+6n+20)(2+n)/81$ & 2 & 10  & 3  \\
2 & $ \includegraphics[angle=0, width=0.1\textwidth]{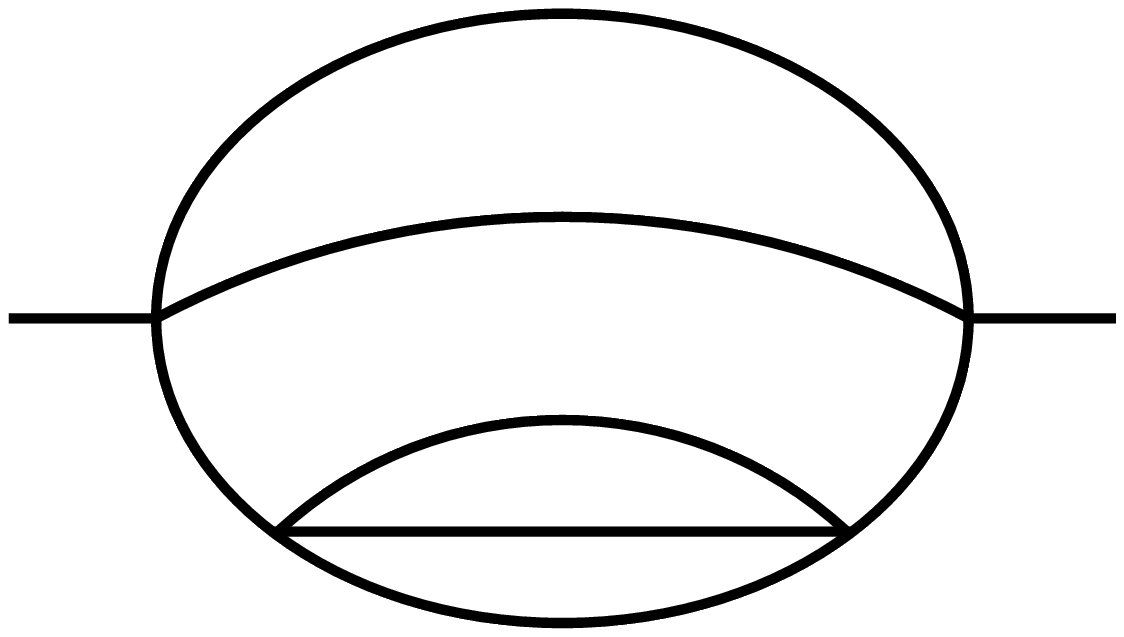}$
    & e112|e3|333||& $(2+n)^2/9$ &2 & 10  & 3  \\ 
3 & $ \includegraphics[angle=0, width=0.1\textwidth]{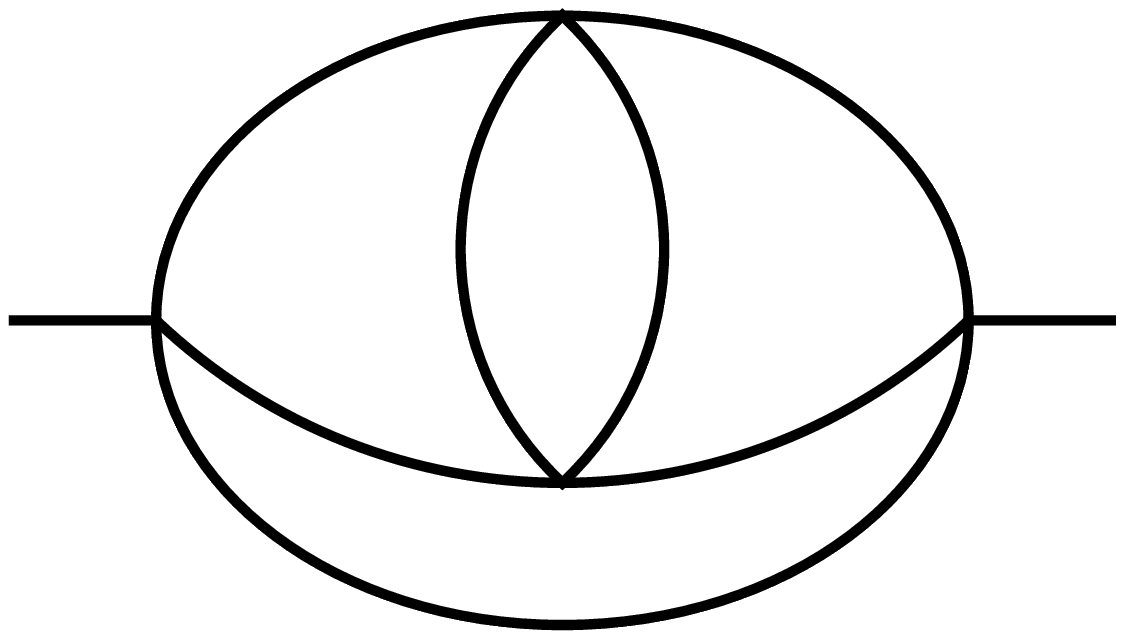}$
    & e123|e23|33||& $(22+5n)(2+n)/81$ &3 & 18  & 4  \\ 
4 & $ \includegraphics[angle=0, width=0.1\textwidth]{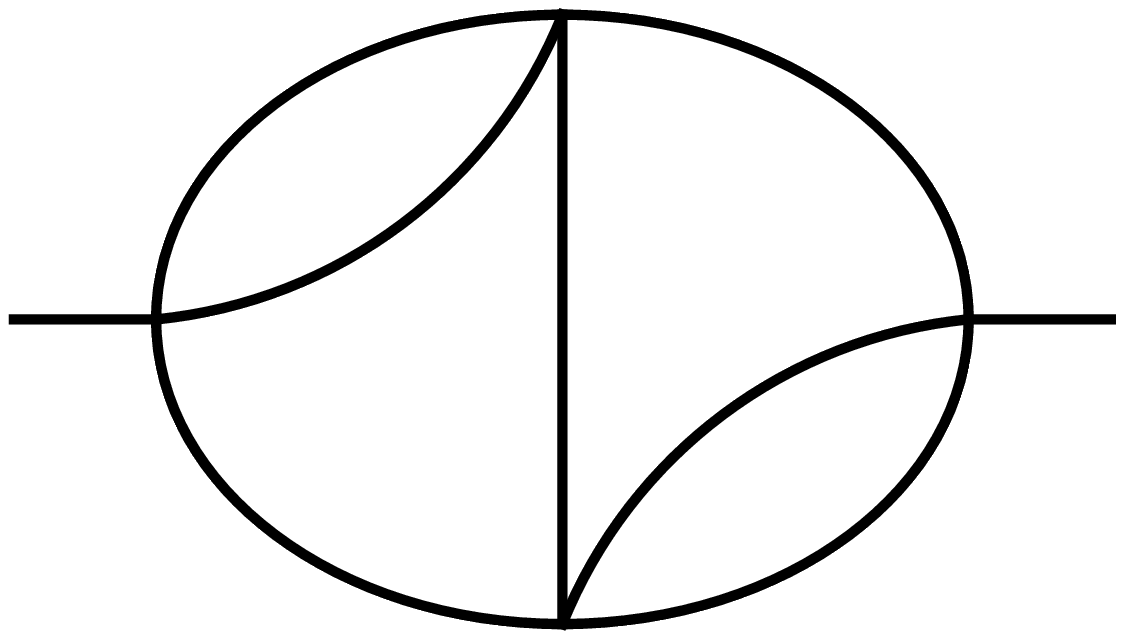}$ 
    & e112|23|33|e|& $(22+5n)(2+n)/81$& 5 & 28  & 7  \\
\multicolumn{4}{r}{Total:} & 12  &  66  & 17 \\ \hline
 \multicolumn{7}{c}{5 loop}   \\ \hline 
1 & $ \includegraphics[angle=0, width=0.1\textwidth]{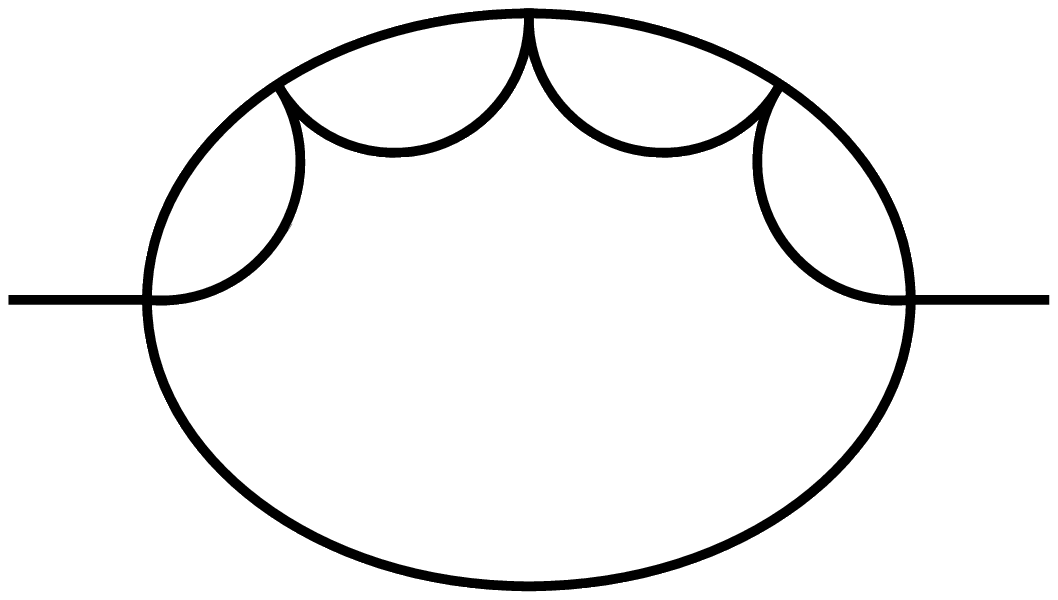}$
    & e112|33|e44|44||& $(n^3+8n^2+24n+48)(2+n)/243$& 2  & 15  & 4  \\ 
2 &$ \includegraphics[angle=0, width=0.1\textwidth]{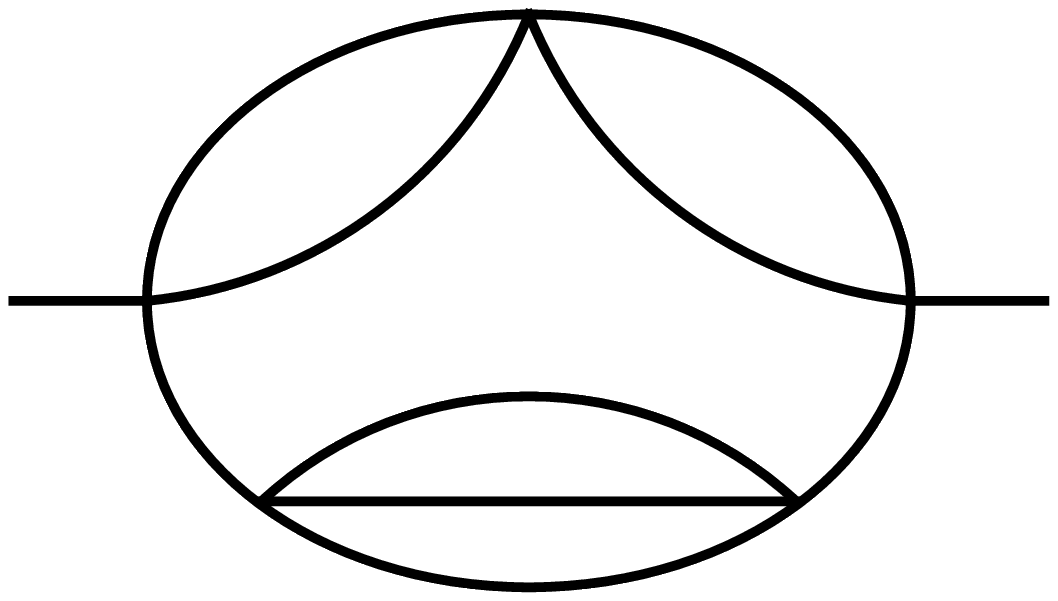}$ & 
    e112|33|444|e4|| & $(8+n)(2+n)^2/81$& 3  &  45  & 9  \\ 
3 &$ \includegraphics[angle=0, width=0.1\textwidth]{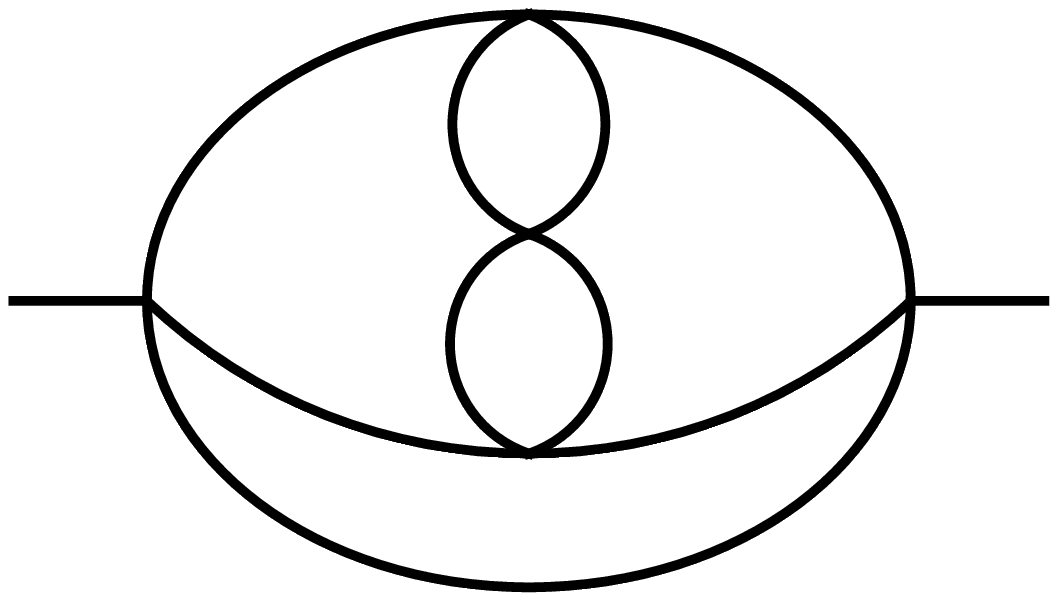}$ &
    e123|e23|44|44|| & $(2+n)(3n^2+22n+56)/243$& 3 &  30 & 8  \\
4 & $ \includegraphics[angle=0, width=0.1\textwidth]{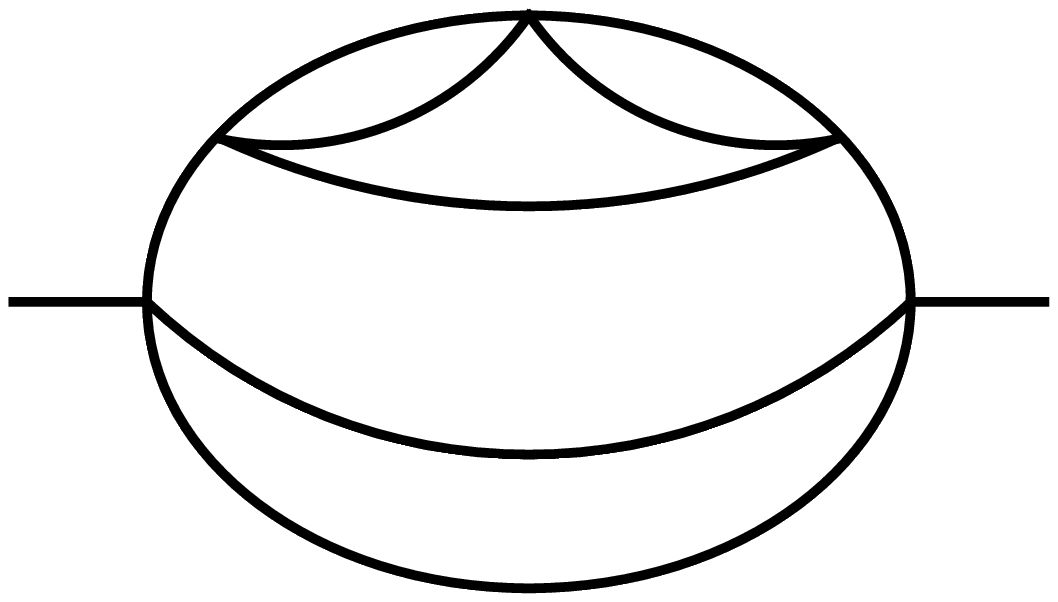}$ & 
    e112|e3|344|44|| &$ (8+n)(2+n)^2/81$ &  4 &  30  & 8  \\ 
5 & $ \includegraphics[angle=0, width=0.1\textwidth]{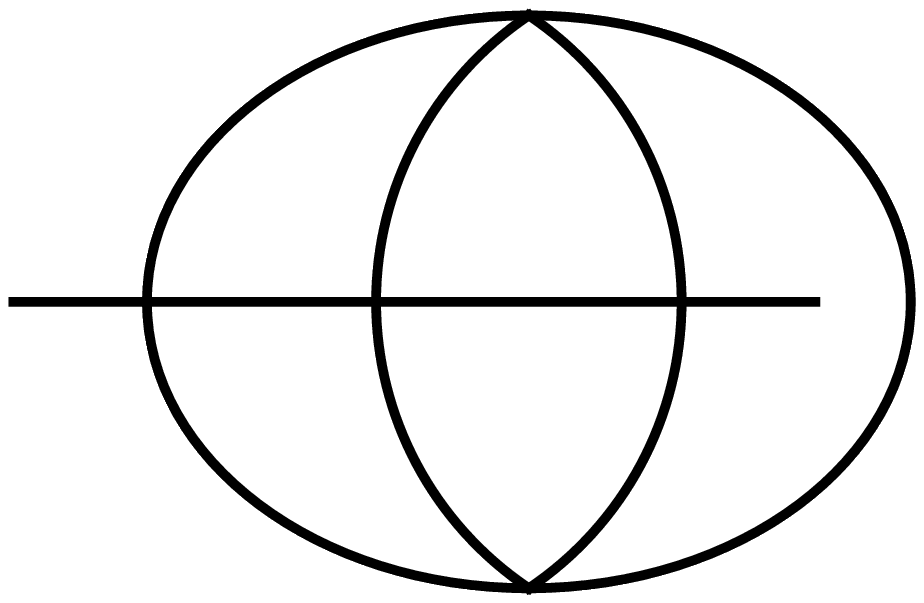}$ & 
    e123|234|34|4|e| & $(22+5n)(2+n)/81$ & 6 &  90  & 10  \\ 
6 & $ \includegraphics[angle=0, width=0.1\textwidth]{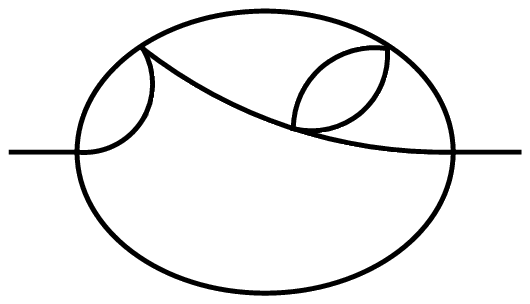}$ & 
    e112|34|e34|44|| & $(2+n)(3n^2+22n+56)/243$& 9 & 90  & 17  \\ 
7 & $ \includegraphics[angle=0, width=0.1\textwidth]{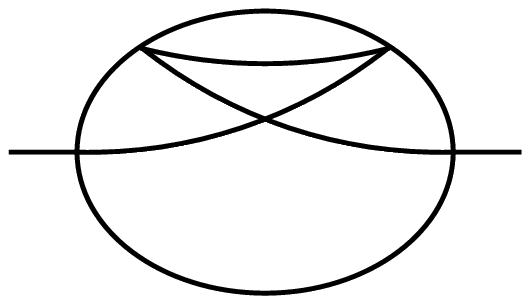}$ &  
    e123|e24|34|44|| & $(n^2+20n+60)(2+n)/243$& 12 & 120  & 23  \\ 
8 & $ \includegraphics[angle=0, width=0.1\textwidth]{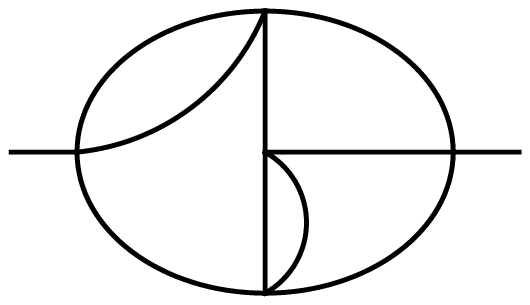}$ & 
    e112|34|334|4|e| & $(n^2+20n+60)(2+n)/243$& 21 &  270  & 51  \\ 
9 & $ \includegraphics[angle=0, width=0.1\textwidth]{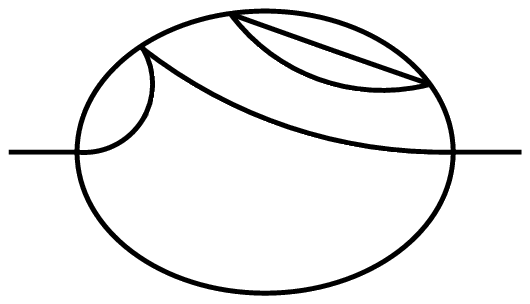}$ & 
    e112|23|e4|444|| & $(8+n)(2+n)^2/81$& 7  & 65  & 19  \\ 
10 & $ \includegraphics[angle=0, width=0.1\textwidth]{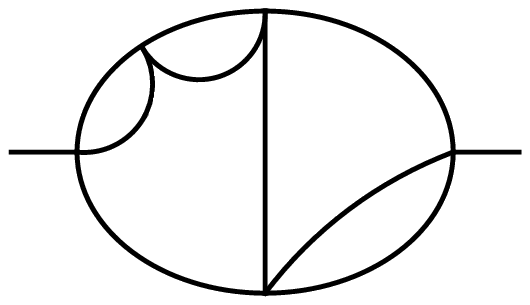}$ &      e112|23|44|e44||  & $(2+n)(3n^2+22n+56)/243$& 13  & 150  & 29  \\ 
11 & $ \includegraphics[angle=0, width=0.1\textwidth]{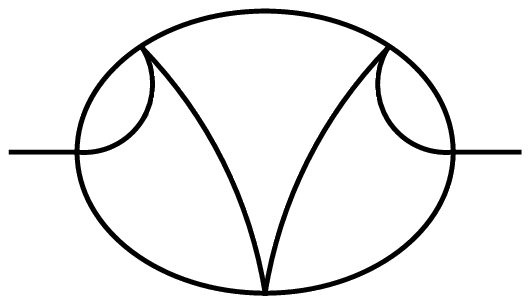}$ & 
    e112|23|34|44|e|  &$(n^2+20n+60)(2+n)/243$ & 12  & 120  & 23  \\ 
 \multicolumn{4}{r}{Total:} & 92  &  1025  & 201 
\end{longtable}

All diagrams are calculated using the Sector Decomposition method~\cite{BINOTH2004375} which was firstly adapted for critical dynamics in Ref.~\cite{AIKV18}. The numerically calculated diagrams are presented in Table~\ref{tab:all_diagrams_exp} in~\ref{Sec_appendix_A}. The sum of the contributions of the diagrams, taking into account the combinatorial factors and the coefficients $a(n)$ shown in Table~\ref{tab:topology}, determines the coefficients $A^{(i)}$ from expansion~\eqref{Gamma}. For clarity, we demonstrate the values of these coefficients in the one-component case $(n=1)$. They read as follows:
\begin{eqnarray}
&&A^{(2)} = 1/8\, \ln{(4/3)} \, \varepsilon^{-1} + 0.0416702(4) + 0.0647910(6)\, \varepsilon + 0.0446045(9)\, \varepsilon^{2}, \nonumber\\ 
&&A^{(3)} =1/8\, \ln{(4/3)} \, \varepsilon^{-2} + 0.1368518(6)\, \varepsilon^{-1} + 0.2709195(13) + 0.285296(3) \varepsilon,   \nonumber\\  
&&A^{(4)}=27/64\, \ln{(4/3)}\, \varepsilon^{-3}+ 0.347358(4) \, \varepsilon^{-2}+ 0.867242(12) \varepsilon^{-1}+ 1.29757(3),  \nonumber\\ 
&&A^{(5)}= 27/40\, \ln{(4/3)}\, \varepsilon^{-4} + 0.77317(4)\, \varepsilon^{-3} + 2.41421(13) \, \varepsilon^{-2} + 4.8300(4)\, \varepsilon^{-1}. \nonumber
\end{eqnarray}
Substituting the coefficients found $A^{(i)}$ into Eq.~\eqref{Gamma} as well as using the expression known from static consideration:
\begin{eqnarray}\label{Zg}
&& Z_g = 1 + u\frac{8+n}{6\varepsilon} +u^2 \left(\frac{(8+n)^2}{36\varepsilon^2}- \frac{14+3n}{24\varepsilon} \right)+ \\
&&\quad + u^3 \left( \frac{(8+n)^3}{216\varepsilon^3}- \frac{7(112 + 38 n + 3 n^2)}{432\varepsilon^2} +\frac{2960 + 922 n + 33 n^2}{5184\varepsilon}+\frac{\zeta(3)(22 + 5 n)}{54\varepsilon}\right)+O(u^4),\nonumber
\end{eqnarray}
we can find the renormalization constant $Z_1$ from the requirement of the elimination of the poles in Eq.~\eqref{Gamma}. For further calculations, we need only the coefficient $Z_1^{(1)}$ at the first-order pole in $\varepsilon$ in $Z_1$. It can be written in the following form:
\begin{eqnarray} \label{z1}
Z_1^{(1)}&=& \frac{n+2}{3} \left[a_{20} u^2+(a_{30}+a_{31}\,n)u^3+(a_{40}+a_{41}\,n+a_{42}\,n^2)u^4+ \right.\nonumber \\
& & \left.+(a_{50}+a_{51}\,n+a_{52}\,n^2+a_{53}\,n^3)u^5\right]\, ,
\end{eqnarray}
where the coefficients $a_{ij}$ are defined in Table~\ref{tab:zn}.
\begin{table}[h!] 
\caption{Numerical values of coefficients from Eq.~\eqref{z1}. It is worth noting that the values of the coefficients at higher powers  in $n$ ($a_{42}$, $a_{53}$) within each of the corresponding order are known exactly from the analysis of the asymptotics when $n$ tends to infinity~\cite{HHM72}.}
\label{tab:zn}
\centering
\renewcommand*{\arraystretch}{1.2}
\renewcommand*{\tabcolsep}{22.4pt}
\begin{tabular}{clcl}
\hline\hline
Coef. & \multicolumn{1}{c}{Value}&Coef.&\multicolumn{1}{c}{Value} \\
\hline
$a_{20}$ &  $-0.0359602590565$                   & $a_{42}$ &\phantom{$-$}$0.0001671773304$ \\ 
$a_{30}$ &  \phantom{$-$}$0.0105255461549$       & $a_{50}$ &\phantom{$-$}$0.04926(29)$      \\  
$a_{31}$ &  \phantom{$-$}$0.0013156932693$       & $a_{51}$ & \phantom{$-$}$0.01357(17)$ \\ 
$a_{40}$ &  $-0.0210863(5)$                      & $a_{52}$ & \phantom{$-$}$0.000386(16)$ \\  
$a_{41}$ &  $-0.00399345(12)$                    & $a_{53}$ & $-0.0000082188688$\\
\hline\hline
\end{tabular}
\end{table}
Let us note also that the two- and three-loop contributions to $Z_1^{(1)}$ are known analytically:
\begin{eqnarray}
Z_1^{(1)} &=& -u^2\frac{(2 + n)}{24} \ln{(4/3)} - u^3\frac{(2 + n)(8 + n) }{27}  \nonumber\\
&& \frac{1}{96}\left(\pi^2 - 8\, \operatorname{Li}_2(1/4) + \ln{(4/3)}\left(-6 - 21 \ln{(3}) + 13 \ln{(4)}\right)\right)+O(u^4)\quad,
\end{eqnarray}
where $\operatorname{Li}_2(x)$ is the dilogarithm. These coefficients were first obtained in the Ref.~\cite{HHM72} and~\cite{AV84}, respectively.

\section{Anomalous dimension and critical exponent z}\label{Sec_expansion}
The RG functions are determined by the renormalization constants. In particular, $Z_1$ determines the anomalous dimension $\gamma_1$ as:
\begin{equation}\label{gamma1}
\gamma_1=\beta(u)\,\partial_u \log Z_1 \, .
\end{equation}
The expansion for the $\beta$-function is currently known up to the sixth~\cite{KP17} and seventh~\cite{S18} loop accuracy. We do not need its explicit form since the above-mentioned connection between the coefficients at higher poles and coefficients at the first pole -- $Z_1^{(1)}$ -- together with Eq.~\eqref{Z1} allow us to present $\gamma_1$ in a simpler manner:
\begin{equation}\label{gamma1a}
\gamma_1=-u\partial_u Z_1^{(1)}.
\end{equation}
The dynamic critical exponent $z$ is expressed, in turn, in terms of the $\gamma_1^*\equiv \gamma_1(u_*) $ function $\gamma_1(u)$ at the fixed point $u_*$ as well as the Fisher exponent $\eta$ by means of the following relation~\cite{V04}:
\begin{equation}\label{z}
z=2+\gamma_1^*-\eta \, .
\end{equation}
The value of the fixed point $u^*$ determined by zero of the $\beta$-function ($\beta(u^*)=0$), and $\eta$ can be taken with the necessary accuracy from the $\phi^4$ field theory:
\begin{eqnarray}
u^*& =& \varepsilon\frac{3}{8 + n}+\varepsilon^2\frac{9(14 + 3 n)}{(8 + n)^3}+ \varepsilon^3\frac{1}{(n+8)^5}(-5912.23 - 2022.99 n - 175.120 n^2  \nonumber \\
&-& 12.375 n^3)+ \varepsilon^4\frac{1}{(n+8)^7}(771918.1 + 411076.2 n + 90771.7 n^2 + 10753.7 n^3\nonumber \\
&+& 438.956 n^4 - 0.3125 n^5),\, \\
\eta &=& \varepsilon^2 \frac{2 + n}{2 (8 + n)^2}- \varepsilon^3\frac{(2 + n) (-272 - 56 n+n^2)}{8 (8 + n)^4} - \nonumber\\
&-&\varepsilon^4\frac{2+n}{ (8 + n)^6}(1096.74 + 334.330 n + 36.9984 n^2 + 7.1875 n^3 + 0.15625 n^4)+\nonumber \\
&+& \varepsilon^5 \frac{(2+n)}{ (8 + n)^8}(222336.8 + 117827.7 n + 25396.1 n^2 + 2572.12 n^3 + 13.8027 n^4  \nonumber \\ 
&-& 4.58183 n^5 + 0.0486946 n^6)\, . 
\end{eqnarray}
As a result, the $\varepsilon$ expansion for the dynamic exponent $z$ reads:
\begin{eqnarray}\label{eqn:arbitrary_n_z_exp}
&& z(\varepsilon, n) = 2 + \nonumber \\
&&  +\varepsilon^2 \frac{(2+n)}{ 2(8 + n)^2}(-1 + 6 \ln{(4/3)} )+\varepsilon^3\frac{(2+n)}{8(8 + n)^4}( 162.462 + 31.9024 n  -1.27352 n^2) \nonumber \\ 
&&  +\varepsilon^4 \frac{(2+n)}{ 32 (8 + n)^6}(-23752.4(16) -6929.0(8) n  -770.28(13) n^2 -170.470(6) n^3  -4.24424 n^4)\nonumber \\
&& + \varepsilon^5 \frac{(2+n)}{ 128 (8 + n)^8}\left(1.986(25)10^{7} + 1.038(23)10^{7} n  + 2.15(8)10^{6} n^2+ 1.92(13)10^{5} n^3 \nonumber \right.\\
&&  \qquad \qquad \qquad \qquad \left. -3.8(10)10^{3} n^4-626(29) n^5+ 7.50053 n^6\right),
\end{eqnarray}
where in the one-component case ($n=1$) it is
\begin{eqnarray}\label{z_one}
 && z(\varepsilon) = 2+ 0.01344616 \, \varepsilon^2 +0.01103628\,\varepsilon^3 - 0.0055791(4)\, \varepsilon^4+0.01773(31)\, \varepsilon^5\, .
\end{eqnarray}
A more detailed notation of the coefficients entering equation~\eqref{eqn:arbitrary_n_z_exp} is  presented in~\ref{Sec_appendix_B}. In the next section we apply different resummation techniques in order to extract proper numerical estimates for the critical dynamic exponent $z$ in the case of different values of order parameter dimensionality.

\section{Borel resummation \label{Sec_resummation}}
During the last fifty years a large variety of resummation methods were implemented in order to get relevant numerical estimates for critical exponents~\cite{GuillouZinn85,kazakov1979,KP17}. The most popular among them use the Borel transformation with consequent analytical continuation by means of conformal mapping of a special form~\cite{K16,BCK16,S18,KTS79,kazakov1979}. This method also has a lot of realizations. Basically, they differ from each other in the strategy of choosing resummation parameters, as well as in different ways of using some additional information about the system behaviour and its observables. 

In this work, we address to two specific implementations~\cite{KP17,AEHIKKZ2021}. Both of them use the knowledge of asymptotic behavior of RG expansions of the A model in higher orders. Let us briefly describe how one can take advantage of this fact.  
Suppose we have the asymptotic expansion 
\begin{equation}\label{eqn:asympto_expansion_general}
  A(\varepsilon)=\sum_{k=0}^\infty A_k \varepsilon^k,
\end{equation}
for which only the first $N+1$ coefficients $A_k$ are known as well as its high-order asymptotic behavior (HOA):
\begin{equation}\label{eqn:hoab_A_expansion}
  A_k \xrightarrow[k \rightarrow \infty]{} c\, k! k^{b_0}(-a)^k \,.
\end{equation}
For the model A these parameters were previously calculated in Ref.~\cite{honkonen}: $a=3/(n+8)$ and $b_{0}=3 + n/2$. For the truncated part of expansion~\eqref{eqn:asympto_expansion_general} one can construct the Borel transformation in the following form: 
\begin{eqnarray}\label{eqn:borel_im} 
&A^N(\varepsilon)=\sum\limits_{k=0}^{N} A_{k}\varepsilon^{k} = \int\limits_0^\infty dt \, e^{-t} t^b \sum\limits_{k=0}^{N} B^b_{k} \, (\varepsilon t)^k = \int\limits_0^\infty dt \, e^{-t} t^b F^{N}_b(\varepsilon t)\, , \quad 	B^b_{k} = \dfrac{(-1)^k A_k}{\Gamma(k+b+1)},&
\end{eqnarray}
where the function $F_b^{N}(x)$ is called the Borel image. In practice, within the RG approach any information about observables, say, critical exponents, can be extracted only from the truncated expansion $A^N(\varepsilon)$. From a formal point of view, however, if the entire series were known, i.e. the whole function $A(\varepsilon)$, then the corresponding Borel image $F_b(x)\equiv F_b^{\infty}(x)$ would possess a finite radius of convergence ($1/a$), which formally does not allow the integration in Eq.~\eqref{eqn:borel_im} to be performed. In order to overcome this problem, it is necessary to find an analytic continuation for the function $F_b(x)$ beyond the circle of convergence and here some ambiguity exists. For example, it can be achieved by analytical continuation of $F_b(x)$ via a conformal mapping of a special form:
\begin{equation}
w(x)=\dfrac{\sqrt{1+ax}-1}{\sqrt{1+ax}+1}, \quad x(w)=\dfrac{4w}{a(1-w)^2}.   
\end{equation}
where the first expression transforms the real semiaxis $x\in (0,\infty)$ into $w\in (0,1)$ making the integration region within a circle of convergence. Here we assume that all singularities of the Borel image lie on the cut $(-\infty,−1/a]$, which are mapped now onto the unit circle $|w|=1$. In terms of $w$, the expansion $A(\varepsilon)$ can be rewritten as:
\begin{equation}\label{eqn:A_n_CM}
 A(\varepsilon)=\int_0^{\infty} dt e^{-t}t^b\left(\frac{\varepsilon t}{w(\varepsilon t)}\right)^\lambda Q(w(\varepsilon t))=\int_0^{\infty} dt e^{-t}t^b G_{b,{\lambda}}(\varepsilon t). 
\end{equation}
As was said above, we are limited only by a finite number of terms for $A(\varepsilon)$; in this situation the approximation for the function $Q(x)$ can be taken as:
\begin{equation}\label{eqn:q_as_w}
Q_N(x)=\sum_{n=0}^N q_k x^k\,.
\end{equation}
The first $N+1$ coefficients $q_k$ are determined from the condition of coincidence of expression~\eqref{eqn:A_n_CM} reexpanded in terms of $\varepsilon$ up no $N+1$ terms with the corresponding coefficients $A_k$ from~\eqref{eqn:borel_im}. The HOA of these coefficients has the same form in Eq.~\eqref{eqn:hoab_A_expansion} with the same $b_0$ if one chooses $b=b_0+3/2$. The exponent $\lambda$ in Eq.~\eqref{eqn:A_n_CM} determines the \textit{strong coupling} asymptotics:
\begin{equation}
A (\varepsilon) \sim \varepsilon^\lambda, \quad     \varepsilon \rightarrow \infty.
\end{equation}
From this point, the  above two methods differ in further modifications designed to improve the convergence of estimates.  

\subsection{KP17}
In Ref.~\cite{KP17}, the following idea was suggested. Apart from the two presented parameters -- $b$ and $\lambda$ -- the third parameter can be introduced to improve the results even further. It enters in the following \textit{homographic transformation}:
\begin{equation}\label{eqn:homographic_trans}
	\varepsilon(\varepsilon')  \rightarrow \dfrac{\varepsilon'}{1+q\varepsilon'}\:, \quad \varepsilon'(\varepsilon)  \rightarrow \dfrac{\varepsilon}{1-q\varepsilon}\;,
\end{equation}
which allows one to reexpand the original $\varepsilon$ expansion as a series in $\varepsilon'$. Further, all the above performed steps will be taken with the new $\varepsilon'$ expansion. Thus, the final expression with which the observables can be estimated is:
\begin{equation}\label{eqn:A_n_CMkp17}
A(\varepsilon)\approx A_{b,\lambda,q}^N(\varepsilon)=\int_0^{\infty} dt e^{-t}t^b G_{b,\lambda,q}^N\left(\dfrac{\varepsilon t}{1-q\varepsilon}\right). 
\end{equation}
It is obvious that the estimates obtained by formula~\eqref{eqn:A_n_CMkp17}, in fact depend on the values of the parameters $b$, $\lambda$, and $q$ and the order of PT ($N$). In Ref.~\cite{KP17}, the strategy of their optimal determination was suggested. In short, they are defined in such a way that in the vicinity of the selected values the estimate of the physical observable was the least sensitive to their variation (for details see Ref.~\cite{KP17}).

\subsection{Free boundary conditions}

In Ref.~\cite{AEHIKKZ2021}, in order to improve the convergence, the authors suggested a technique called the \textit{free boundary condition} which is inspired by the resummation method suggested earlier for calculating static exponents in Ref.~\cite{guida1998critical}. The authors of~\cite{guida1998critical} proposed to use the known values of critical exponents in spatial dimensions (further called the \textit{boundary dimensions}) different from the one (further called the \textit{dimension of consideration}) in which these quantities are desired. However, the information of this kind is not always known. It is this situation that is observed in the case of the critical dynamic exponent $z$ for an arbitrary value of the order parameter dimensionality $n$. In order to overcome this problem, the authors of Ref.~\cite{AEHIKKZ2021} used the trick according to which the value of the critical exponent in boundary spatial dimension should be considered as a variable quantity, a certain value of which is chosen based on the best convergence of estimates of the critical exponent in the considered dimension with growing PT order. Formally, this procedure can be described as follows. Let us have an initial series $A(\varepsilon)$, then the first thing we should do is its reexpanding taking into account the assumed (actually varied) value $A_{\textup{b}}$ in the boundary spatial dimension (4-$\varepsilon_{\textup{b}}$): 
\begin{equation}
 A(\varepsilon)=A_{\textup{b}}+(\varepsilon_{\textup{b}}-\varepsilon)B_{\textup{b}}(\varepsilon), \quad B_{\textup{b}}=\sum\limits_{k=0}^{\infty}B_{\textup{b},k}\varepsilon^k,
\end{equation}
where by construction the HOAs from~\eqref{eqn:hoab_A_expansion} for the initial series $A(\varepsilon)$ and the reexpanded one $B_{\textup{b}}(\varepsilon)$ coincide with each other.
\begin{figure}
\center
\includegraphics[width=1\textwidth]{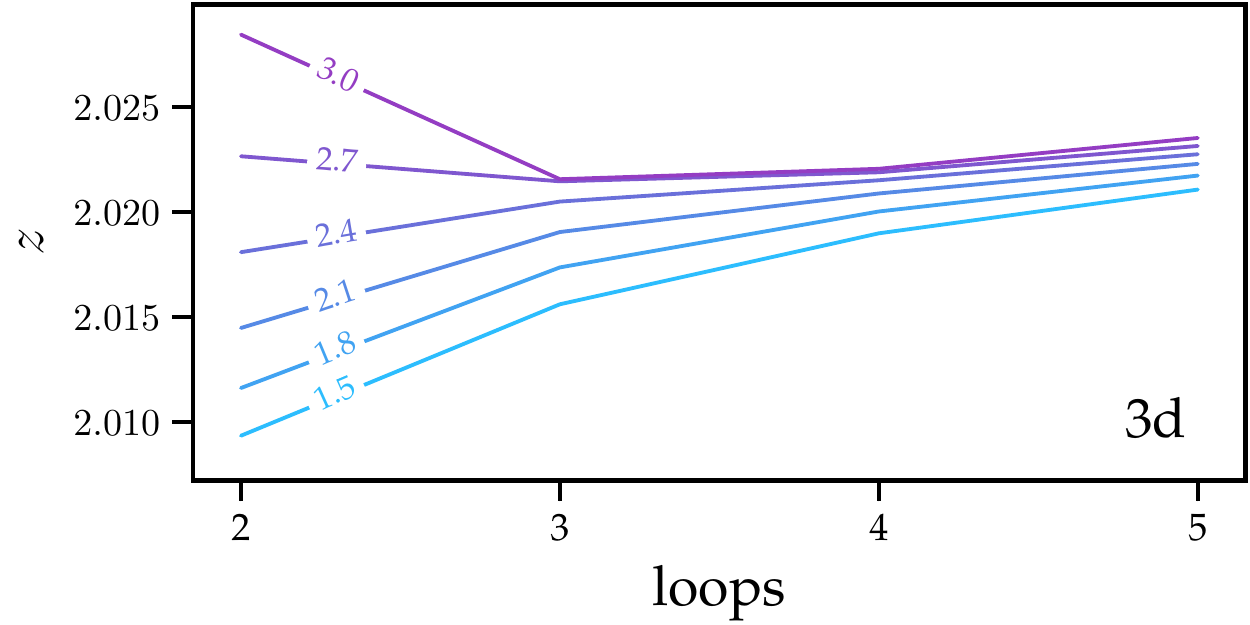} 
\caption{The behavior of estimates for the critical exponent $z$ for $n=1$ based on the order of PT for different values of $\lambda$.}
\label{fig:cb_wot_bc}
\end{figure}
As noted earlier, in practice we have only a limited number of terms -- $N+1$ -- for the series $A(\varepsilon)$, which leads to a limited number (the same $N+1$) of known terms for the new expansion $B_{\textup{b}}(\varepsilon)$. Now for this truncated expansion $B_{\textup{b}}^N(\varepsilon)$ we repeat the general steps with the Borel transformation and conformal mapping~\eqref{eqn:A_n_CM} but without using the homographic transformation from KP17~\eqref{eqn:homographic_trans}. Thus, the following expression can be considered as the final formula for calculation of the exponent $z$ within the present section:  \begin{equation}\label{eqn:free_boundary_final_expression}
 A(\varepsilon)\approx A_{\textup{b},\lambda,b}^N(\varepsilon)=A_{\textup{b}}+(\varepsilon_{\textup{b}}-\varepsilon)B_{\textup{b},\lambda,b}^N(\varepsilon).
\end{equation}

Here, in contrast to KP17, the value of $b$ is fixed and chosen according to the HOA for the A model; only the value of the critical dynamic exponent in the boundary spatial dimension $z_{\textup{b}}$ as well as the strong coupling parameter $\lambda$ will be variable. 
\begin{figure}
\label{fig:n=1}\includegraphics[width=0.49\textwidth]{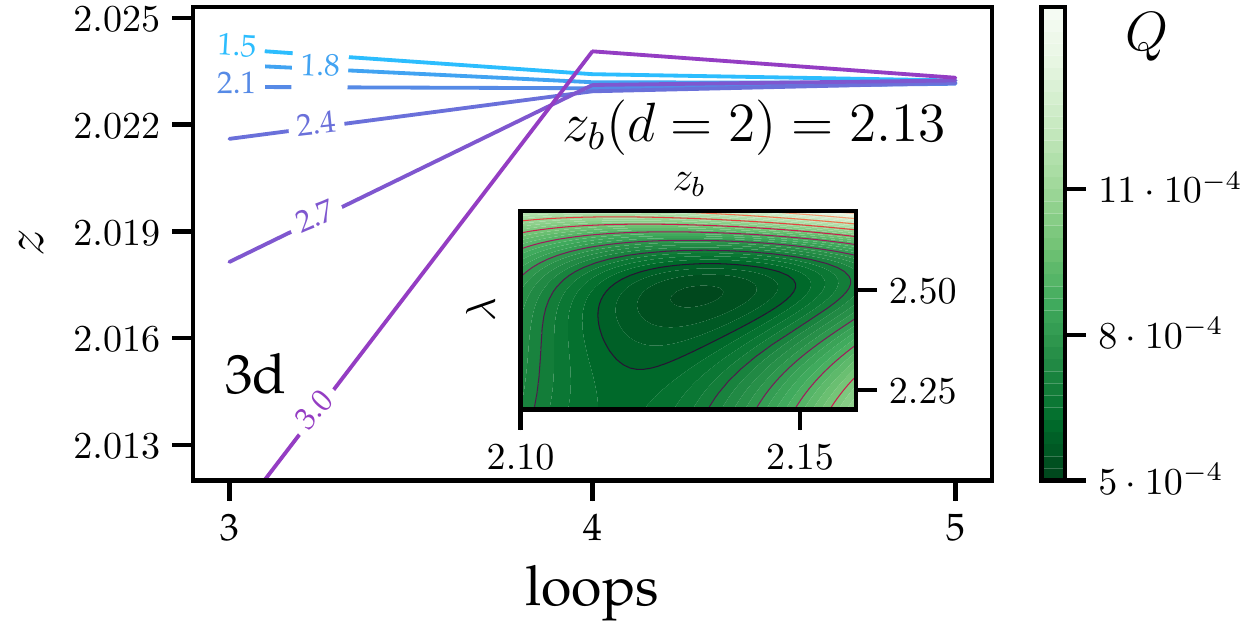}
\label{fig:n=2}\includegraphics[width=0.49\textwidth]{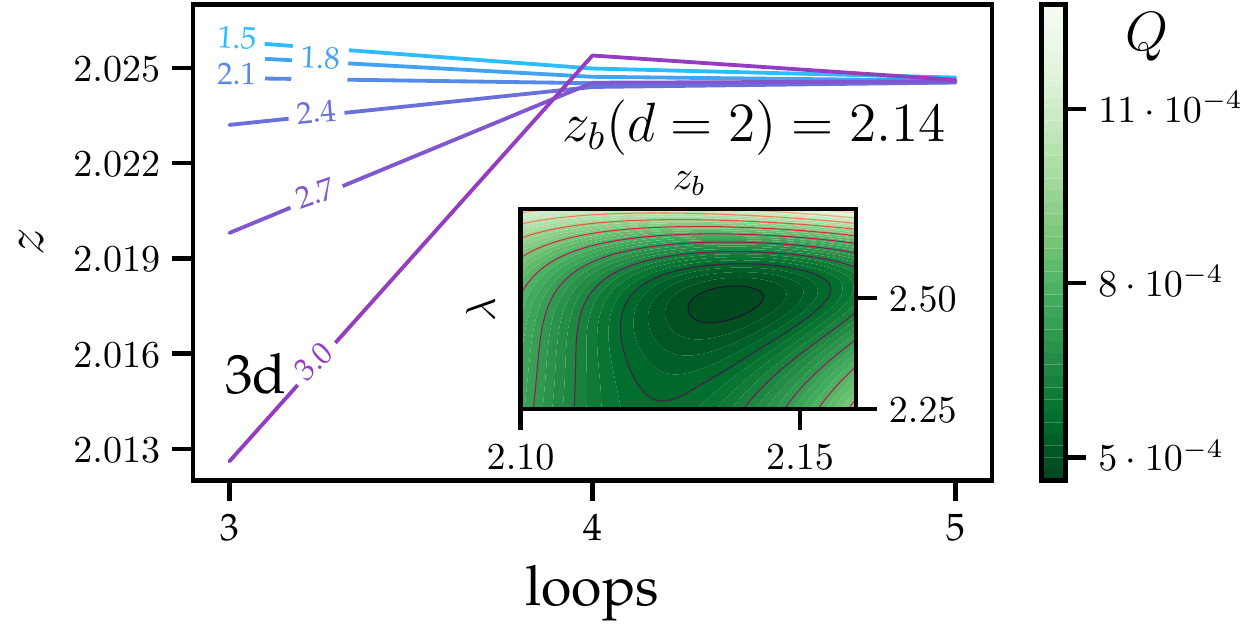}\\
\label{fig:n=5}\includegraphics[width=0.49\textwidth]{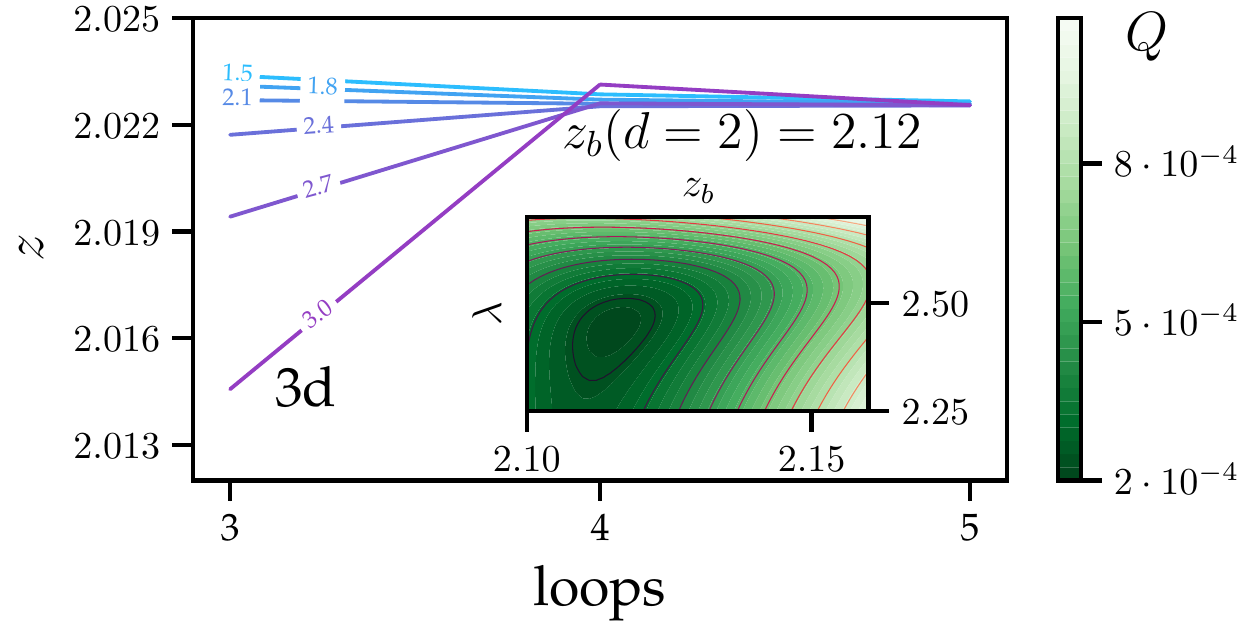}
\label{fig:n=10}\includegraphics[width=0.49\textwidth]{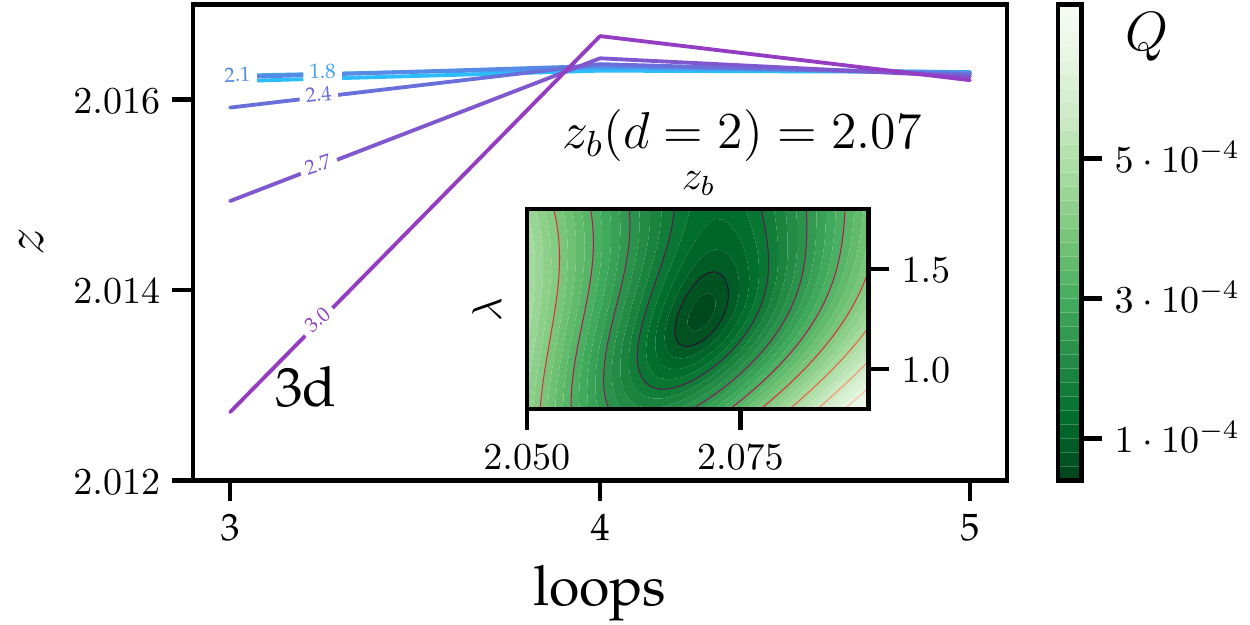}
\caption{The behaviour of numerical estimates for critical exponent $z$ based on the order of PT for different values of $\lambda$ in the case of taking into account the boundary condition $z_{2d}=2.13$ for $n=1$ (upper left), $z_{2d}=2.14$ for $n=2$ (upper right), $z_{2d}=2.12$ for $n=5$ (lower left), and $z_{2d}=2.07$ for $n=10$ (lower right).}
\label{fig:Q_criterion}
\end{figure}

For clarity, we have demonstrated the behavior of $z$ with an increase in the number of loops when the value of $z_{\textup{b}}$ at the boundary dimension ($d=2$) is not fixed (Fig.~\ref{fig:cb_wot_bc}) and taken equal to different numbers for different values of the order parameter dimension $n$ (Fig.~\ref{fig:Q_criterion}). Each line corresponds to its own value for the $\lambda$ parameter. The dependencies in Fig.~\ref{fig:Q_criterion} show the trend of convergence of the calculation results to a specific value as the number of the considered terms increases, similar behavior takes place in the case of the exactly solvable zero-dimensional model~\cite{K16}. It should be also noted that this trend is much more pronounced than in the case of Fig.~\ref{fig:cb_wot_bc}, where parameter $z_\textup{b}$ is not used. As the optimal values of $\lambda$ and $z_b$, one should choose those that provide the best convergence to a given limit value based on a smaller number of terms of PT under consideration.
It is clear from Fig.~\ref{fig:Q_criterion} that in order to determine these optimal values, one can require that the slope of the straight line would decrease in the fastest way with the growth of PT. In order to write down some formal convergence criterion, we can address to the equation of the straight line passing through the points of four- ($z^{(4)}$) and five-loop ($z^{(5)}$) approximations as: 
\begin{eqnarray}\label{eqn:slope_eq} 
z(l)=a(\lambda)(l-5)+b(\lambda)\,,
\end{eqnarray}
where $l$ is number of loops, and $a(\lambda)=z^{(5)}-z^{(4)}$ and $b(\lambda)=z^{(5)}$. Hence, in order to achieve the highest density of the straight lines in~\eqref{eqn:slope_eq}, we demand the least sensitivity of the slope of the straight line $z(l)$ to the $\lambda$ variation that, in turn, compels the derivatives $\partial_{\lambda}a$, $\partial_{\lambda}b$, and $ \partial_{\lambda}^2b $ to take their minimum values. In this regard, it is natural to consider as a criterion of convergence the requirement of the minimum value of the following quantity:
\begin{equation}\label{eqn:Q}
Q=\sqrt{\left(\partial_{\lambda}a\right)^{2}+\left(\partial_{\lambda}b\right)^{2}+\left(\partial_{\lambda}^2 b\right)^{2}}.
\end{equation}
\begin{table}[b!] 
\caption{ \label{tab:z_extimates_d_n} Numerical estimates of the critical dynamic exponent $z$ in the case of different spatial dimensions $d$ and values of order parameter dimensionality $n$. The estimates are found by means of the KP17 and FBC resummation procedures of the expansion~\eqref{eqn:arbitrary_n_z_exp}. At each number the set of optimal values of the resummation parameters is presented. The  abbreviation WA denotes the weighted average, which is constructed based on the five loop KP17 and FBC results and is taken as the final estimate for each $\bm{\eucal{A}}^{d,n}$ model.}
\begin{center}
\renewcommand*{\arraystretch}{1.1}
\renewcommand*{\tabcolsep}{12.23pt}
\begin{tabular}{rllll}
\hline\hline
$n$ & Dim. &\multicolumn{1}{c}{Free b.c.} & \multicolumn{1}{c}{KP17} & \multicolumn{1}{c}{W. A.}\\
\hline
\multirow{2}{*}{$-1$}
& d=2 &  $2.06(2)_{[1.78, 0.00038]}$ & $2.08(2)_{[8.0,2.46,0.22]}$    & $2.070(14)$  \\
& d=3 &  $2.0123(10)_{[1.74,0.00034]}$  & $2.0134(9)_{[8.0,2.46,0.22]}$ & $2.0127(8)$\\
\hline
\multirow{2}{*}{$0$}
& d=2 & $2.11(3)_{[2.48, 0.00051]}$ & $2.13(3)_{[8.25,2.62,0.22]}$    & $2.12(2)$  \\
& d=3 & $2.020(2)_{[2.48,0.00051]}$  & $2.021(2)_{[8.5,2.42,0.24]}$ & $2.0205(11)$\\
\hline
\multirow{2}{*}{$1$}
& d=2 &  $2.13(4)_{[2.49,0.00052]}$ & $2.15(3)_{[9.25,2.42,0.24]}$    & $2.14(2)$  \\
& d=3 &  $2.023(2)_{[2.49,0.00052]}$  & $2.0239(14)_{[9.25,2.42,0.24]}$ & $2.0236(8)$\\
\hline
\multirow{2}{*}{$2$}
& d=2 &  $2.14(4)_{[2.49,0.00047]}$ & $2.16(3)_{[9.5,2.66,0.22]}$    & $2.15(2)$  \\
& d=3 &  $2.024(3)_{[2.49,0.00047]}$  & $2.0249(13)_{[10.0,2.46,0.22]}$ & $2.0246(10)$\\
\hline
\multirow{2}{*}{$3$}
& d=2 &  $2.13(3)_{[2.49,0.00039]22}$ & $2.15(2)_{[10.0,2.68,0.2]}$    & $2.145(15)$  \\
& d=3 &  $2.024(2)_{[2.49,0.00039]}$  & $2.0247(12)_{[10.75,2.44,0.2]}$ & $2.0244(8)$\\
\hline
\multirow{2}{*}{$4$}
& d=2 &  $2.13(3)_{[2.48,0.00030]}$ & $2.13(3)_{[10.5,2.62,0.16]}$    & $2.130(15)$  \\
& d=3 &  $2.024(2)_{[2.48,0.00030]}$  & $2.0237(10)_{[11.0,2.5,0.16]}$ & $2.0238(7)$\\
\hline
\multirow{2}{*}{$5$}
& d=2 &  $2.12(2)_{[2.45, 0.00021]}$ & $2.12(2)_{[10.5,2.64,0.12]}$    & $2.120(10)$  \\
& d=3 &  $2.0223(14)_{[2.45, 0.00021]}$  & $2.0225(8)_{[10.75,2.5,0.12]}$ & $2.0224(5)$\\
\hline
\multirow{2}{*}{$6$}
& d=2 &  $2.104(12)_{[2.39,0.00013]}$ & $2.107(13)_{[10.5,2.46,0.08]}$    & $2.105(6)$  \\
& d=3 &  $2.0210(10)_{[2.39,0.00013]}$  & $2.0218(13)_{[2.75,1.0,0.42]}$ & $2.0213(7)$\\
\hline
\multirow{2}{*}{$7$}
& d=2 &  $2.094(7)_{[2.31,0.00006]}$ & $2.0946(10)_{[9.75,2.7,0.02]}$    & $2.0945(9)$  \\
& d=3 &  $2.0197(5)_{[2.31,0.00006]}$  & $2.0202(8)_{[2.5,1.0,0.34]}$ & $2.0199(4)$\\
\hline
\multirow{2}{*}{$8$}
& d=2 &  $2.085(3)_{[1.90,0.000008]}$ & $2.085(9)_{[8.25,2.34,0.0]}$    & $2.085(2)$  \\
& d=3 &  $2.0185(2)_{[1.90,0.000008]}$  & $2.0187(5)_{[2.5,1.2,0.22]}$ & $2.0186(2)$\\
\hline
\multirow{2}{*}{$9$}
& d=2 &  $2.077(3)_{[1.48,0.00004]}$ & $2.079(5)_{[2.25,1.32,0.12]}$    & $2.078(2)$  \\
& d=3 &  $2.0174(2)_{[1.48,0.00004]}$  & $2.0174(2)_{[2.25,1.32,0.12]}$ & $2.01740(10)$\\
\hline
\multirow{2}{*}{$10$}
& d=2 &  $2.070(6)_{[1.28,0.00005]}$ & $2.068(3)_{[2.0,1.54,0.0]}$    & $2.069(2)$  \\
& d=3 &  $2.0163(5)_{[1.28,0.00005]}$  & $2.0162(2)_{[1.5,1.38,0.02]}$ & $2.01622(13)$\\
\hline
\hline
\end{tabular}
\end{center}
\end{table}
Thus, the minimum of $Q$ from~\eqref{eqn:Q} will be found in the case of different spatial dimensions and the number of order parameter components. In the insets in Fig.~\ref{fig:Q_criterion} we demonstrate the typical behaviour of $Q$ in the vicinity of its minimal value as a projection to the ($z_{\textup{b}},\lambda$)-surface, which corresponds to the best convergence of estimates for the 3d when the boundary dimension is $d=2$ in case of different values of order parameter dimensionality.

\section{Numerical results}
\begin{figure}
\center
\includegraphics[width=0.87\textwidth]{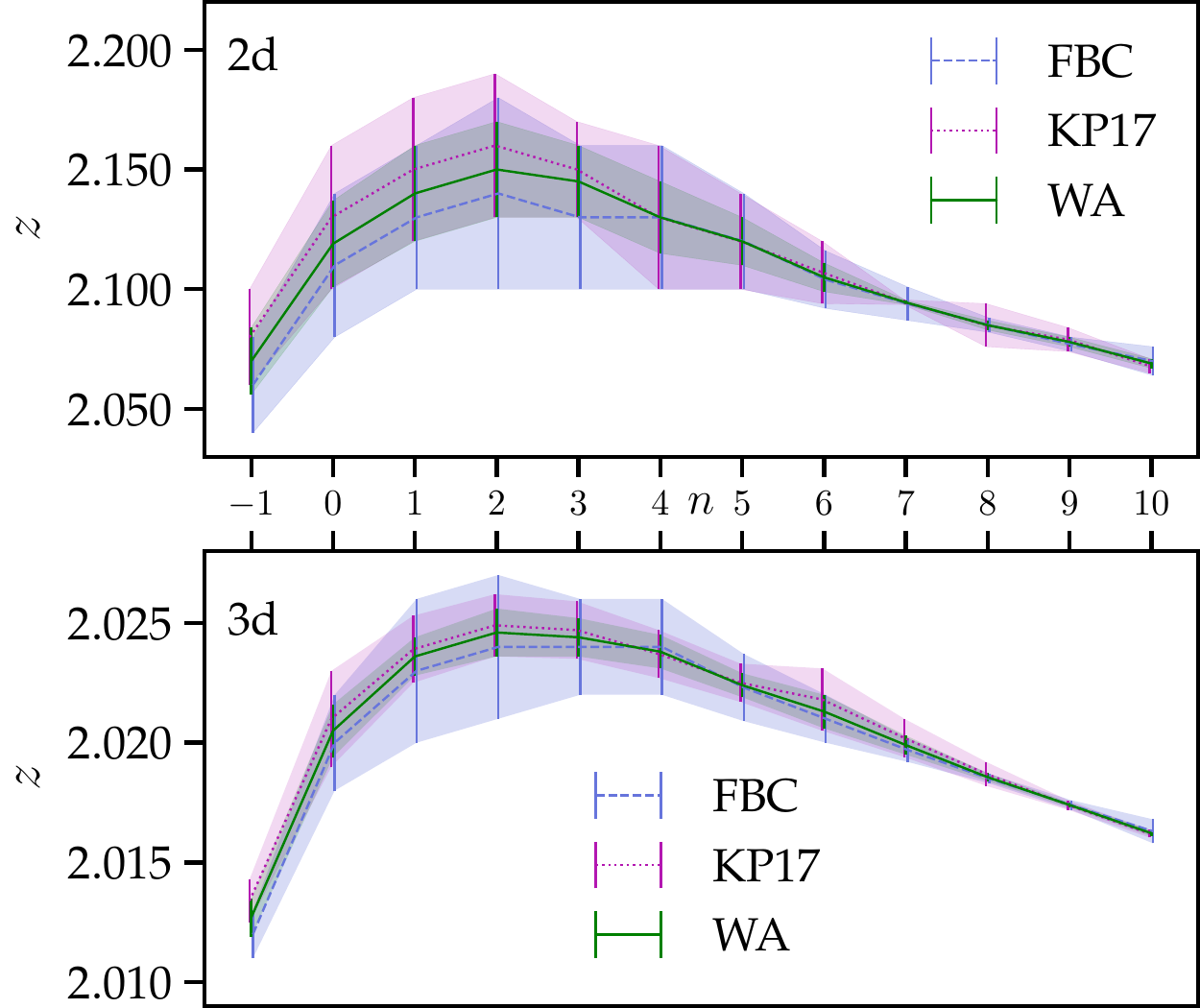} 
\caption{The behavior of estimates for the critical exponent $z$ for 2d and 3d cases with increasing order parameter dimensionality $n$ calculated on the basis of the FBC and KP17 methods.}
\label{fig:trends}
\end{figure}

Following two resummation strategies described in the previous section, based on~\eqref{eqn:arbitrary_n_z_exp} we extract 2d and 3d numerical estimates for the dynamic critical exponent $z$ for different $n$. These numbers are presented in Table~\ref{tab:z_extimates_d_n}. The error of each of the estimates, in addition to the internal error of the particular resummation procedure, takes into account the error of the coefficients of the 4th and 5th PT orders for the $\varepsilon$ expansion of the exponent $z$, i.e. the estimates of the critical exponent $z$ were also calculated for the series with shifted values of the coefficients, and the obtained "forks" were taken into account. As a final numerical estimates for $z$ in the case of specific $n$ and d, we have resorted to the weighted average (WA) in accordance with the procedure described in Ref.~\cite{PhysRevD.103.116024}, where the inverse error bars were taken as weights. For greater clarity, in Fig.~\ref{fig:trends} we depicted the trends of the behavior of the critical exponent $z$ with an increase in the value of the order parameter dimension $n$.

As we can see, all the numbers obtained using the two resummation techniques are in complete agreement with each other for all values of the order parameter dimension $n$ taking into account the corresponding error bars. Note also that in the case of three dimensions, the agreement between each of the methods is much better than in the case of two dimensions and the amplitude of errors is lower. This, of course, is not surprising, due to large values of the expansion parameter ($\varepsilon=2$). In three dimensions FBC gives much more conservative estimates of errors in contrast to KP17. As we can see in both dimensions, the maximum value of the exponent is reached in the case of the two-component A model. Further decrease is estimates can be easily explained, at least due to the general damping of the expansion coefficients with an increase in the dimension of the order parameter.

\section{Summary \label{Sec_conclusion}}

Thus, in this work, we have calculated the five-loop RG expansions for the $n$-component A model of critical dynamics within the Minimal Subtraction scheme. The possibility of obtaining expansions of this length is dictated by the fact that the authors managed to decrease the number of actually relevant diagrams using the \textit{diagram reduction technique} and to apply the effective methods for calculating each diagram by means of the Sector Decomposition technique that had previously been widely used in critical statics. To obtain the high-precision numerical estimates for the critical exponent $z$, that in fact has the prime of physical importance, we applied various resummation techniques for asymptotic series, which were proposed by the authors of this work. Due to the possibility to change the number of the order parameter components, as well as to analyze the critical dynamic exponent $z$ in various spatial dimensions, in this work we have given an exhaustive description of the behavior of the critical dynamic exponent of the model A which describes a wide range of physical systems. For example, for the most interesting physical values of the order parameter dimension ($n=1$), our three-dimensional estimate ($2.0236(8)$) is in good agreement with the most accurate results that have been obtained using alternative approaches($2.0245(15)$~\cite{hasenbusch2020} and $2.0237(55)$~\cite{krinitsyn2006calculations}).

\section*{Acknowledgment}

We would like to thank R.Guida for the helpful discussion and sharing his notes. The work of A.K. was supported by Grant of the Russian Science Foundation No 21-72-00108. The work of M.H. was supported by the Ministry of Education, Science, Research and Sport of the Slovak Republic(VEGA Grant No. 1/0535/21). We are grateful to the Joint Institute for Nuclear Research for allowing us to use their supercomputer “Govorun”.

\newpage

\appendix

\section*{Appendix}

\section{Diagrams for $Z_1$ up to five-loop contributions \label{Sec_appendix_A}}
In Table~\ref{tab:all_diagrams_exp} we present numerical values for all diagrams of the renormalization constant $Z_1$ up to five-loop contributions as part of the Laurent expansion in terms of $\varepsilon$. The coefficients of expansions are given analytically where it is possible, otherwise numerical values are suggested.
\renewcommand*{\arraystretch}{1.3}
\renewcommand*{\tabcolsep}{2.1pt}
\begin{longtable}{cll}
\caption[Numerical values of all diagrams for the renormalization constant $Z_1$ up to five loops in $\varepsilon$. The way of describing of diagram topology is chosen in accordance with the Nickel notation.]{Numerical values of all diagrams for renormalization constant $Z_1$ up to the five loops in $\varepsilon$. The way of description of diagram topology is chosen in accordance with Nickel notation.} \label{tab:all_diagrams_exp} \\
\hline
\hline
\multicolumn{1}{c}{N.} &
\multicolumn{1}{c}{Topology} &
\multicolumn{1}{c}{Numerical results}  \\
\hline
\endfirsthead
\multicolumn{3}{c}%
{\tablename\ \thetable{} diagrams} \\
\hline
\multicolumn{1}{c}{N.} &
\multicolumn{1}{c}{Topology} &
\multicolumn{1}{c}{Numerical results}  \\
\hline
\endhead
\hline \multicolumn{3}{r}{{Continuation on the next page}} \\
\endfoot
\hline \hline
\endlastfoot
\multicolumn{3}{c}{2 loop}   \\
\hline
1 & e111|e|&  $ 1/8 \ln{(4/3)}\varepsilon^{-1} + 0.041670207(4) + 0.0647910(6) \varepsilon + 0.0446045(9) \varepsilon $  \\ \hline
\multicolumn{3}{c}{3 loop}   \\ \hline 
1 & e112|22|e|& $1/4 \ln{(4/3)} \varepsilon^{-2} + 0.1368518(6) \varepsilon^{-1} + 0.27091943(4) + 0.285296(3)\varepsilon $  \\ \hline
\multicolumn{3}{c}{4 loop}   \\ \hline 
1 & e112|33|e33||& $ 9/64 \ln{(4/3)}\varepsilon^{-3} +  0.1070793(11) \varepsilon^{-2} + 0.256086(3) \varepsilon^{-1} + 0.358251(6) $  \\ 
2 & e112|e3|333||& $-0.000852(3) \varepsilon^{-2} + -0.005031(10) \varepsilon^{-1} + -0.017487(15)$    \\ 
3 & e123|e23|33||& $3/32 \ln{(4/3)}\varepsilon^-3 + 0.0848713(9) \varepsilon^{-2} + 0.216134(3) \varepsilon^{-1} + 0.377071(5) $   \\ 
4 & e112|23|33|e|& $3/16 \ln{(4/3)} \varepsilon^{-3} + 0.1562589(12) \varepsilon^{-2} + 0.400053(4) \varepsilon^{-1} + 0.579729(8) $   \\ \hline
\multicolumn{3}{c}{5 loop}   \\ 
\hline
1 & e112|33|e44|44||& $3/40 \, \ln{(4/3)} \,\varepsilon^{-4} +0.073163(4) \varepsilon^{-3} +0.204788(16) \varepsilon^{-2} + 0.35477(5) \varepsilon^{-1} $    \\ 
2 & e112|33|444|e4|| & $ -0.000666(12) \varepsilon^{-3} -0.00451(4) \varepsilon^{-2} -0.01797(12)\varepsilon^{-1} $    \\ 
3 & e123|e23|44|44|| &  $1/40 \, \ln{(4/3)} \,\varepsilon^{-4} + 0.031582(3)\varepsilon^{-3} +  0.098616(9)\varepsilon^{-2} $   \\ 
 & & $+ 0.23615(3) \varepsilon^{-1} $\\ 
4 & e112|e3|344|44|| &  $ -0.00133(3) \varepsilon^{-3}-0.01160(9) \varepsilon^{-2}-0.0524(3) \varepsilon^{-1} $    \\ 
5 & e123|234|34|4|e| &  $0.0172897(9)\varepsilon^{-2} + 0.076055(4) \varepsilon^{-1} $   \\ 
6 & e112|34|e34|44|| &  $1 /10 \, \ln{(4/3)}\, \varepsilon^{-4} + 0.111931(5) \varepsilon^{-3} + 0.33142(2)\varepsilon^{-2} + 0.67233(6) \varepsilon^{-1} $    \\ 
7 & e123|e24|34|44|| &  $1/20 \, \ln{(4/3)}\, \varepsilon^{-4} +  0.070358(4) \varepsilon^{-3} + 0.23961(2)\varepsilon^{-2} + 0.61230(6) \varepsilon^{-1}$    \\ 8 & e112|34|334|4|e| &  $ 3/20 \, \ln{(4/3)} \, \varepsilon^{-4} + 0.182275(8)\varepsilon^{-3} + 0.58732(3)\varepsilon^{-2} + 1.21353(9)\varepsilon^{-1}$   \\ 
9 & e112|23|e4|444|| & $ -0.00196(2)\varepsilon^{-3} -0.01250(7) \varepsilon^{-2} -0.0491(3) \varepsilon^{-1}$   \\ 
10 & e112|23|44|e44||  & $7/40 \, \ln{(4/3)}\, \varepsilon^{-4} + 0.188694(8)\varepsilon^{-3} + 0.57368(3) \varepsilon^{-2} + 1.03326(9) \varepsilon^{-1} $   \\ 
11 & e112|23|34|44|e|  & $1/10 \, \ln{(4/3)} \, \varepsilon^{-4} + 0.119119(6) \varepsilon^{-3} + 0.39009(2) \varepsilon^{-2} + 0.75114(8) \varepsilon^{-1} $  \\ \hline
\end{longtable}

\section{Five-loop $\varepsilon$ expansion for critical exponent $z$ \label{Sec_appendix_B}}
In this Appendix we present the five-loop $\varepsilon$ expansion for the critical dynamic exponent $z$ for an arbitrary value of the order parameter dimension $n$. In terms of the coefficients $b_{ij}$ from Table~\ref{tab:coefficients_b} the expansion can be written as:
\begin{eqnarray}\label{eqn:z_exp_b_coef}
 z(\varepsilon, n) = 2 &+&\varepsilon^2 \frac{(2+n)}{ 2(8 + n)^2}(b_{20})+\varepsilon^3\frac{(2+n)}{8(8 + n)^4}( b_{30}+b_{31}n+b_{32}n^2) \nonumber \\ 
&+&\varepsilon^4 \frac{(2+n)}{ 32 (8 + n)^6}(b_{40}+b_{41}n+b_{42}n^2+b_{43}n^3+b_{44}n^4)+\nonumber \\
&+&\varepsilon^5 \frac{(2+n)}{ 128 (8 + n)^8}\left(b_{50}+b_{51}n+b_{52}n^2+b_{53}n^3+b_{54}n^4+b_{55}n^5+b_{56}n^6 \right).\qquad
\end{eqnarray}
\renewcommand*{\arraystretch}{1.3}
\renewcommand*{\tabcolsep}{6.9pt}
\begin{longtable}{ll}
\caption[Analytical values of the coefficients $b_{ij}$ entering the Eq.~\eqref{eqn:z_exp_b_coef} in terms of the variables $a_{ij}$ from Table~.]{Analytical values of coefficients $b_{ij}$ entering Eq.~\eqref{eqn:z_exp_b_coef} in terms of variables $a_{ij}$ from Table~\ref{tab:zn}.} 
\label{tab:coefficients_b} \\
\hline
\hline
\multicolumn{1}{c}{Coef.} &
\multicolumn{1}{c}{Value} \\
\hline
\endfirsthead
\multicolumn{2}{c}%
{\tablename\ \thetable{} Values of expansion coefficients.} \\
\hline
\multicolumn{1}{c}{Coef.} &
\multicolumn{1}{c}{Value} \\
\hline
\endhead
\hline \multicolumn{2}{r}{{Continuation on the next page}} \\
\endfoot
\hline \hline
\endlastfoot
\hline
 $b_{20}$ & $ -1 + 6 \ln{(4/3)}$ \\ \hline 
 $b_{30}$ & $-272 - 16128 \, a20 - 13824\, a_{30} $ \\ \hline 
 $b_{31}$ & $-56 - 3456\, a_{20} - 1728 \,a_{30} - 13824 \,a_{31} $ \\ \hline 
 $b_{32}$ & $1 -1728\, a_{31}$ \\  \hline
 $b_{40}$ & $-46144 - 2227200\, a_{20} - 6967296 \,a_{30} - 
 3538944 \,a_{40} + 67584 \zeta(3) + 
 3244032\, a_{20} \zeta(3) $ \\  \hline
 $b_{41}$ & $-17920 - 918528 \,a_{20} - 2363904 \,a_{30} - 6967296 \,a_{31} - 884736 \, a_{40} - 3538944\, a_{41}$ \\
 & $  + 23808 \zeta(3) + 1142784\, a_{20} \zeta (3)$ \\ \hline 
 $b_{42}$ & $-1124 - 83328 \,a_{20} - 186624 \,a_{30} - 2363904 \,a_{31} - 55296 \,a_{40} - 884736 \,a_{41}$ \\
 & $- 3538944 \,a_{42} + 1920 \zeta(3) + 92160 \,a_{20} \zeta(3) $\\ \hline 
 $b_{43}$ & $230 + 6336 \,a_{20} - 186624\, a_{31} - 55296\, a_{41} - 
  884736 \,a_{42} $ \\ \hline 
 $b_{44}$ & $5 - 55296 \,a_{42}$ \\  \hline
 $b_{50}$ & $-5655552 - 214384640 \,a_{20} - 1547403264 \,a_{30} - 2378170368 \,a_{40} - 
 849346560 \,a_{50} $\\
 & $+ (720896 \pi^4)/5 +(23068672 \,a_{20} \pi^4)/5 + 
 21921792 \zeta (3) + 1436024832 \,a_{20} \zeta (3) $ \\
 & $-60948480 \zeta(5) + 1401421824 \,a_{30} \zeta (3) - 2925527040 \,a_{20} \zeta(5)$ \\ \hline 
 $b_{51}$ & $-2912768 - 127770624 \, a_{20} - 841052160 \,a_{30} - 1547403264 \,a_{31} - 
 1104150528 \, a_{40}$ \\
 & $- 2378170368 \, a_{41}- 849346560 \,a_{51} + (
 434176 \pi^4)/5 $ \\
 & $+ (13893632 \,a_{20} \pi^4)/5 + 11530240 \zeta(3) + 
 793313280 \,a_{20} \zeta(3) + 1401421824 \,a_{31} \zeta(3) $ \\
 &$- 33259520 \zeta(5) - 1596456960 \,a_{20} \zeta(5) - 318504960 \, a_{50}+ 668860416 \,a_{30} \zeta(3) $\\ \hline 
 $b_{52}$ & $-(787544/3) - 22469632 \,a_{20} - 143824896 \, a_{30} - 841052160 \,a_{31} - 
 164560896 \,a_{40} $\\
 & $- 1104150528 \,a_{41}-39813120 \,a_{50} - 
 318504960 \,a_{51} - 849346560 \, a_{52} $\\
 & $+ (95232 \pi^4)/5 + (
 3047424 \,a_{20} \pi^4)/5 + 1244160 \zeta(3) + 111329280 \, a_{20} \zeta(3) $ \\
 &$+  101523456 \,a_{30} \zeta(3) + 668860416 \, a_{31} \zeta(3) - 6113280 \zeta(5) - 
 293437440 \,a_{20} \zeta(5)$\\ 
 &$- 2378170368 \,a_{42} $\\ \hline
 $b_{53}$ & $121472 + 1018880 \,a_{20} - 5121792 \, a_{30} - 143824896 \, a_{31} - 7962624 \,a_{40}$\\
 & $ - 
 164560896 \,a_{41} - 1104150528 \, a_{42} - 1658880 \, a_{50} - 39813120 \, a_{51} - 318504960 \,a_{52} $\\
 & $- 849346560 \,a_{53} + (9088 \pi^4)/5 - 137984 \zeta(3) - 1867776 a_{20} \zeta(3) $\\
 &$+ 
 4976640 \, a_{30 }\zeta(3) + 101523456 \, a_{31} \zeta(3) - 445440 \zeta(5)+ (
 290816 \, a_{20 }\pi^4)/5$\\
  &$- 21381120 \,a_{20} \zeta(5)$\\ \hline
  $b_{54}$ & $27620 + 569856 \,a_{20} + 342144 \,a_{30 }- 5121792 \, a_{31} - 7962624 \, a_{41}$\\
 & $- 
 164560896 \,a_{42} - 1658880 \,a_{51} - 318504960 \, a_{53} + 
 64 \pi^4 + 2048 \,a_{20} \pi^4 $ \\
 &$- 20800 \zeta(3) - 774144 a_{20} \zeta(3) + 
 4976640 a_{31} \zeta(3) - 10240 \zeta(5) - 39813120 \,a_{52 }$\\ \hline 
 &$- 491520 a_{20} \zeta(5)$\\ \hline 
 $b_{55}$ & $2798/3 + 640 \, a_{20} + 342144 \, a_{31} - 7962624 \,a_{42} - 1658880 \,a_{52}$\\
 &$-39813120 \,a_{53} - 288 \zeta(3)$\\ \hline
 $b_{56}$ & $13 - 1658880 a_{53} - 16 \zeta(3)$ \\ 
\end{longtable}

\noindent Using the coefficients from Table~\ref{tab:coefficients_b} we come to the equations~\eqref{eqn:arbitrary_n_z_exp} for $z$ for an arbitrary value of $n$.

\bibliographystyle{unsrt}
\bibliography{modelA_cor}
\end{document}